\documentclass[11pt,a4paper]{article}
\usepackage{amssymb, amsmath,framed,mathrsfs}
\usepackage[colorlinks]{hyperref}
\usepackage{diagrams}
\usepackage[margin=1in]{geometry}
\newcommand{\rem}[1]{}
\newcommand{\de}{{\rm d}}
\newcommand{\PH}{{\mathbf{P}_{\!}\mathscr{H}}}
\newcommand{\R}{{{Re}}}
\newcommand{\I}{{{Im}}}
\newcommand{\Tr}{{\operatorname{Tr}}}
\newcommand{\Sch}{\text{Schr\"odinger}\,}

\newcommand{\bz}{{\mathbf{z}}}
\newcommand{\bx}{{\mathbf{x}}}
\newcommand{\bp}{{\mathbf{p}}}

\newcommand{\bzeta}{\boldsymbol{\zeta}}

\newtheorem{theorem}{Theorem}

\newtheorem{lemma}[theorem]{Lemma}
\newtheorem{remark}{Remark}

\begin{document}

\title{Geometry and symmetry of quantum\\and classical-quantum variational principles}
\author{Esther Bonet Luz, Cesare Tronci\\
\it\footnotesize Department of Mathematics, University of Surrey, Guildford GU2 7XH, United Kingdom\\
}
\date{}

\maketitle

\begin{abstract}
This paper presents the geometric setting of quantum variational principles and extends it to comprise the interaction between classical and quantum degrees of freedom. Euler-Poincar\'e reduction theory is applied to the Schr\"odinger, Heisenberg and Wigner-Moyal dynamics of pure states. This construction leads to new variational principles for the description of mixed quantum states. The corresponding momentum map properties are presented as they arise from the underlying unitary symmetries. Finally, certain semidirect-product group structures are shown to produce new variational principles for Dirac's interaction picture and the equations of hybrid classical-quantum dynamics.
\end{abstract}

\bigskip

\tableofcontents


\newpage

\section{Introduction}

After Kibble's investigation \cite{kibble1979geometrization} of the geometric properties of quantum state spaces, geometric formulations of quantum dynamics have been attracting much attention over the last decades \cite{Anandan91,AnAh90,AshAb95,BoCaGra91,BroHu01,ChiMel2012,Chrus94,CleMar08,FaKuMaMa10,Grig92,Montgomery91,SaHuKu2011,KhaBroGla01,uhlmann2009geometry}. In turn, geometric quantum dynamics has opened several modern perspectives: for example, Fubini-Study geodesics have been introduced in Grover's quantum search algorithms \cite{MiWa01,Zhao2012} and in time-optimal quantum control \cite{CarHoKoOku06,CarHoKoOku07,Dalessandro01}, while the holonomy features arising from the quantum geometric phase \cite{AhAn87,BoCaGra91} and its non-Abelian extensions \cite{Anandan88non,Chrus94} have been proposed in quantum computation algorithms  \cite{GunWaNa2014,Lucarelli2005,TaNaHa2005}.

Most approaches deal with pure quantum states and involve the geometry of the Hopf fibration
\begin{align*}
S(\mathscr{H})&\to \mathbf{P}_{\!}\mathscr{H}
\\
\psi&\mapsto\psi\psi^\dagger\,,
\end{align*}
where $S(\mathscr{H})$ denotes the unit sphere in a complex Hilbert space $\mathscr{H}$ (so that $\psi\in S(\mathscr{H})$ is a unit vector in $\mathscr{H}$), while $ \mathbf{P}_{\!}\mathscr{H}$ is the corresponding projective space containing the projections $\rho_\psi:=\psi\psi^\dagger$. The geometry of the above Hopf bundle is well known and has been widely studied in the finite dimensional case $\mathscr{H}=\Bbb{C}^n$, although some studies extend to consider infinite-dimensional Hilbert spaces \cite{ChiMel2012}. In the finite dimensional case, one can emphasize the symmetry properties of the Hopf bundle by writing
\begin{equation}\label{bundles}
S(\Bbb{C}^n)=\mathcal{U}(n)/\mathcal{U}(n-1)
\,,\qquad
\mathbf{P}\Bbb{C}^n=S(\Bbb{C}^n)/\mathcal{U}(1)=\mathcal{U}(n)/\big(\mathcal{U}(n-1)\times \mathcal{U}(1)\big)
\,.
\end{equation}
The interplay between the geometry of the Hopf bundle and its symmetries is the basis of geometric quantum dynamics. For example, the emergence of principal bundles leads to the usual horizontal-vertical decomposition in terms of a principal connection that is strictly related to Berry's geometric phase. 
This is a beautiful picture, whose symplectic Hamiltonian properties have been widely investigated after Kibble's work \cite{kibble1979geometrization}. 

In this paper, we aim to present how  this geometric framework emerges naturally from the unitary symmetry properties of quantum variational principles. Time-dependent variational approaches have been most successful in chemical physics (here, we recall the celebrated Car-Parrinello model in molecular dynamics \cite{CarPa85}). The most fundamental quantum variational principle is probably due to Dirac and Frenkel (DF) \cite{dirac1930note,frenkel1934wave}. This action principle produces Schr\"odinger equation $i\hbar\dot\psi=H\psi$ as the Euler-Lagrange equation associated to
\[
\delta\int_{t_1}^{t_2\!}\!\big\langle\psi,i\hbar\dot\psi-{H}\psi\big\rangle\,\de t=0
\,,
\]
where $H$ is the quantum Hamiltonian operator and we introduce the pairing $\langle A,B\rangle$ and the inner product
\[
\langle A|B\rangle=\operatorname{Tr}(A^\dagger B)
\,,\qquad\text{ so that } \qquad
\langle A,B\rangle:=\R\,\langle A|B\rangle
\]
and $\I\,\langle A|B\rangle=\langle iA,B\rangle$. 
Various properties of the variational principle above have been studied over the decades \cite{LowMuk72,kramer1981geometry,Ohta2000}, after it was first proposed in the context of Hartree-Fock mean field theories. For example, it is known that the DF action principle is simply the quantum correspondent of the classical Hamilton's principle on phase space $\delta\!\int_{t_1}^{t_2}\!\big(\boldsymbol{p}\cdot\dot{\boldsymbol{q}}-{H(\boldsymbol{q,p})}\big)\,\de t=0$, so that $\hbar\langle\psi,i\de\psi\rangle$ acquires the meaning of canonical one form on $\mathscr{H}$ (analogously, $\boldsymbol{p}\cdot\de{\boldsymbol{q}}$ is the canonical one form in classical mechanics). However, to our knowledge, an investigation of the geometric symmetry properties of quantum variational principles has not been carried out. Although an early attempt was proposed in \cite{kramer1981geometry}, the emergence of the Hopf bundle in this context has not been presented so far. For example, the momentum maps associated to quantum variational principles have never been considered in the literature, while they are essential geometric features often associated to fundamental physical quantities. Even in the simplest situation, the phase invariance of quantum Lagrangians produces the momentum map identifying the total quantum probability $\|\psi\|^2$.

The present work applies well known techniques in the theory of geometric mechanics \cite{marsden1999introduction}, which focus on the symmetry properties of the dynamics. For example, momentum map structures are seen to emerge: some are new, while others are related to the principal connections associated to quantum geometric phases. Within geometric mechanics, we shall be using the specific tool of Euler-Poincar\'e theory \cite{holm1998euler} that typically applies to variational principles with symmetry. As we shall see, besides recovering well known relations, these theory allows us to formulate new variational principles for various quantum descriptions, such as the  Liouville-Von Neumann equation, Heisenberg dynamics, Moyal-Wigner formulation on phase space and the Ehrenfest theorem for the evolution of expectation values. Some of these descriptions of quantum mechanics have been lacking a variational structure, which is now provided in this paper for the first time. Here, we shall not dwell upon various complications that may emerge in infinite dimensional Hilbert spaces $\mathscr{H}$ and we assume convergence where necessary. When convenient, we shall consider dynamics on finite dimensional spaces and rely on the possibility of extending the results to the infinite dimensional case.

\section{Euler-Poincar\'e variational principles in the Schr\"odinger picture\label{SEC:EPSchr}}
This section presents the Euler-Poincar\'e formulation of quantum dynamics in the \Sch picture. Two main examples are considered: the  Schr\"odinger equation as it arises from the Dirac-Frenkel theory and the Fubini-Study geodesics. Their geometric features will be analyzed in terms of momentum maps.

\subsection{Euler-Poincar\'e reduction for pure quantum states}
Upon denoting by $T\!\mathscr{H}$ the tangent bundle of the Hilbert space $\mathscr{H}$, consider a generic Lagrangian
\begin{equation}\label{genlagr}
L:T\!\mathscr{H}\to\Bbb{R}\,,\qquad
L=L(\psi,\dot\psi)\,,
\end{equation}
so that the assumption of quantum evolution restricts $\psi$ to evolve under the action of that unitary group $\mathcal{U}(\mathscr{H})$, that is
\begin{equation}\label{psievol}
\psi(t)=U(t)\psi_0\,,\qquad
U(t)\in\mathcal{U}(\mathscr{H})
\end{equation}
where $\psi_0$ is some initial condition, whose normalization is ordinarily chosen such that $\|\psi_0\|^2=1$. Then, $\psi_0\in S(\mathscr{H})$ implies $\psi(t)\in S(\mathscr{H})$ at all times. 

The relation \eqref{psievol} takes the Lagrangian $L(\psi,\dot\psi)$ to a Lagrangian of the type $L_{\psi_0}(U,\dot{U})$, which then produces Euler-Lagrange equations for the Lagrangian coordinate $U\in\mathcal{U}(\mathscr{H})$. Moreover, by following Euler-Poincar\'e theory \cite{holm1998euler}, one denotes by $\mathfrak{u}(\mathscr{H})$ the Lie algebra of skew Hermitian operators and defines
\[
\xi(t):=\dot{U}(t)U^{-1}(t)\in\mathfrak{u}(\mathscr{H})\,.
\]
Since $\dot\psi=\xi\psi$, one obtains the reduced Lagrangian
\[
\ell: \mathfrak{u}(\mathscr{H})\times\mathscr{H}\to\Bbb{R}
\,,\qquad
\ell(\xi,\psi):=L(\psi,\xi\psi)
\]
and the Euler-Poincar\'e variational principle
\begin{equation}
\delta\int_{t_1}^{t_2\!}\!\ell(\xi,\psi)\,\de t=0
\,.
\label{varprinc1}
\end{equation}
Then, upon computing
\begin{equation}  \label{variations} 
\delta\xi=\dot\eta+[\eta,\xi]\,,\qquad
\delta\psi=\eta\psi
\end{equation}
where $\eta:=(\delta U)U^{-1}$, one obtains the following result.
\begin{theorem}
Consider the variational principle \eqref{varprinc1} with the auxiliary equation $\dot\psi=\xi\psi$ and the variations \eqref{variations}, where $\eta$ is arbitrary and vanishes at the endpoints. This variational principle is equivalent to the equations of motion
\begin{equation}
\frac{\de}{\de t}\frac{\delta \ell}{\delta\xi} - \left[ \xi,\frac{\delta \ell}{\delta \xi} \right] =  \frac{1}{2} \left( \frac{\delta \ell}{\delta \psi}\psi^{\dagger}-\psi \frac{\delta \ell}{\delta \psi}^{\!\dagger} \right)  \label{EQ:EPpsi} 
,\qquad\qquad
\frac{\de\psi}{\de t} = \xi\psi 
\,.
\end{equation}
\end{theorem}
Here, we use the ordinary definition of variational derivative
\[
\delta F(q) :=\left\langle\frac{\delta F}{\delta q},\delta q\right\rangle
,
\]
for any function(al) $F\in C^{\infty\!}(M)$ on the manifold $M$.
In typical situations, the reduced Lagrangian is quadratic in $\psi$, so that the $\mathcal{U}(1)$-invariance under phase transformations takes the dynamics to the projective space $\PH$. Indeed, as we shall see, the reduced Lagrangian $\ell(\xi,\psi)$ can be written typically in terms of the projection $\rho_\psi=\psi\psi^\dagger$ to produce a new Lagrangian
\[
l: \mathfrak{u}(\mathscr{H})\times\PH\to\Bbb{R}
\,,\qquad
l(\xi,\rho_\psi)=\ell(\xi,\psi)
\,.
\]
In this case, a direct calculation shows that
\begin{equation}\label{variations2}
\delta\rho_\psi=[\eta,\rho_\psi]\,,\qquad
\dot{\rho}_\psi=[\xi,\rho_\psi]
\end{equation}
and the previous theorem specializes as follows
\begin{theorem}
Consider the variational principle $
\delta\!\int_{t_1}^{t_2\!}l(\xi,\rho_\psi)\,\de t=0
$
 with the relations \eqref{variations2} and $\delta\xi=\dot\eta+[\eta,\xi]$, where $\eta$ is arbitrary and vanishes at the endpoints. This variational principle is equivalent to the equations of motion
\begin{equation}
\frac{\de}{\de t}\frac{\delta l}{\delta\xi} - \left[ \xi,\frac{\delta l}{\delta \xi} \right]    =\left[ \frac{\delta l}{\delta\rho_\psi},\rho_\psi \right]  \label{EQ:EPrho} 
,\qquad\quad
\dot{\rho}_\psi=[\xi,\rho_\psi]\,.
\end{equation}
\end{theorem}
Then, the unitary symmetry properties of the Lagrangian naturally take the evolution to the correct quantum state space (for pure states), that is the projective space $\PH$. In the following sections, we shall specialize this construction to two particular examples and we shall present the momentum map properties of the underlying geometry as well as their relation to the usual principal connections appearing in the literature.

\subsection{Dirac-Frenkel variational principle}
It is easy to see that upon following the construction from the previous section, the DF Lagrangian 
\begin{equation}\label{DFLAGR}
L(\psi,\dot\psi)=\big\langle\psi,i\hbar\dot\psi-{H}\psi\big\rangle
\end{equation} 
produces the Euler-Poincar\'e variational principle
\[
\delta\int_{t_1}^{t_2\!}\!\left\langle\psi,i\hbar\xi\psi-{H}\psi\right\rangle\de t=0
\] 
For simplicity, here we are considering a time-independent Hamiltonian operator $H$. Then, upon computing 
\begin{align*}
    \frac{\delta l}{\delta\psi} = 2(i\hbar\xi-H) \psi,~~~~~~~~ \frac{\delta l}{\delta\xi} = -i\hbar \psi \psi^{\dagger} \,,
\end{align*}
the first of \eqref{EQ:EPpsi} yields
\begin{equation}
	[\left(i\hbar\xi-H\right),\psi\psi^{\dagger}] = 0\,.
	 \label{EQ:SCHRU_EP1}
\end{equation}
Upon setting $\mathscr{H}=\Bbb{C}^n$ and making use of the anticommutator bracket $\{A,B\}=AB+BA$, the solution of the above equation can be written as
\begin{equation}
\xi+i\hbar^{-1\!}H=\{\boldsymbol{1}-2\psi\psi^{\dagger},\kappa\}
\label{EQ:projsch}
\end{equation}
for an arbitrary time-dependent skew Hermitian matrix $\kappa(t)$. The meaning of this solution will be clear in Section \ref{sec:momaps}. In the \Sch picture, the  relation \eqref{EQ:projsch} recovers the usual phase arbitrariness, as it is shown by simply using the second in \eqref{EQ:EPpsi} to write
\begin{equation}
i\hbar\dot\psi=H\psi+\alpha\psi\,,
\label{EQ:projsch2}
\end{equation}
where $\alpha(t):=2\hbar\langle i\rho_\psi,\kappa\rangle$ (notice that we have chosen a unit initial vector so that $\|\psi_0\|^2=\|\psi\|^2=1$). 
In the above equation, the   term $\alpha\psi$ generates an arbitrary phase factor. Then, equation \eqref{EQ:projsch2} can be easily written in the form of a projective \Sch equation \cite{kibble1979geometrization}
\[
(1-\psi\psi^\dagger)(i\hbar\dot\psi-H\psi)=0\,,
\] 
which can be recovered from the constrained DF Lagrangian \cite{Ohta2000}
\begin{equation}
L(\psi,\dot\psi,\lambda,\dot{\lambda})=\big\langle\psi,i\hbar\dot\psi-{H}\psi\big\rangle+\lambda(\|\psi\|^2-1)\,.
\label{EQ:constrLagr}
\end{equation}
As we shall see, the right hand side of \eqref{EQ:projsch} modifies the usual Heisenberg picture dynamics.

\subsection{Mixed states dynamics and its Wigner-Moyal formulation}
It is easy to see that all the phase terms in the previous section are consistently projected out by simply defining the Lagrangian $l(\xi,\rho_\psi)=\left\langle\rho_\psi,i\hbar\xi-{H}\right\rangle$ so that the first of \eqref{EQ:EPrho} reads $[(i\hbar\xi-H),\rho_\psi] = 0$ and the second recovers the quantum Liouville equation for pure states. 

In the remainder of this section we shall generalise the previous approach to consider a new variational principle for \emph{mixed quantum states}. Let us consider the Lagrangian
\begin{equation}
l(\xi,\rho)=\left\langle\rho,i\hbar\xi-{H}\right\rangle
\label{mixedlagr}
\end{equation}
where $\xi=\dot{U}U^{-1}$ as before, while $\rho$ is  a density matrix undergoing unitary evolution $\rho(t)=U\rho_0 U^\dagger$. In the case of mixed states, we have $\rho^2\neq\rho$ although the trace invariants $\Tr(\rho^n)$ are still preserved. We notice that a simple computation yields 
\begin{equation}
\delta\rho=[\eta,\rho]\,,\qquad\dot{\rho}=[\xi,\rho]\,,
\label{mixedvariations}
\end{equation}
 and therefore the application of Euler-Poincar\'e theory is straightforward. Then, one obtains precisely the same equations as in \eqref{EQ:EPrho} (upon replacing $\rho_\psi$ by $\rho$), which in turn give
\begin{equation}
[(i\hbar\xi-H),\rho] = 0\,,
\qquad
\dot\rho=[\xi,\rho]
\,.
\label{mixed}
\end{equation}
At this point the Liouville-Von Neumann equation
\[
    i\hbar\dot{\rho} =\left[H,\rho\right] 
\]
is obtained by direct substitution. Notice that the solution of the first equation in \eqref{mixed} differs from \eqref{EQ:projsch},  since $\rho^2\neq\rho$. For example, one has particular solutions of the form $i\hbar\xi-H=\sum_n\alpha_n\rho^n$. This reflects the very different geometric structures underlying mixed states and pure states. For a geometric description of mixed states in terms of coadjoint orbits and orthogonal frame bundles, we refer the reader to \cite{Montgomery91}. We emphasize that the action principle associated to the Lagrangian \eqref{mixedlagr} is very different from the one proposed in  \cite{Heller76} and to our knowledge it has not appeared before.

Motivated by applications in chemical physics, we show how the above variational principle recovers the celebrated Wigner-Moyal picture of quantum dynamics on phase space. This formulation \cite{moyal1949quantum,wigner1932quantum} is based on the Weyl correspondence between linear operators  and phase space functions (see e.g. \cite{Zachos2005}). For simplicity, this section presents the Euler-Poincar\'e formulation on the two-dimensional phase space (one spatial dimension), however this can be easily generalized to higher dimensions.

Consider an arbitrary linear  operator $A\in L(\mathscr{H})$: the corresponding  phase-space  function is given by the Wigner transform $a(x,p)=\mathcal{W}(A)$ and the latter can be inverted by using the Weyl transform, $A=\mathcal{W}^{-1}(a)$. More explicitly, one has
\begin{align*}
    \mathcal{W}(A) &\!:= \frac{1}{\pi\hbar} \int\! \de x^{\prime} \langle x+x^{\prime}|A |x-x^{\prime}\rangle e^{-\frac{2ipx^{\prime}}{\hbar}} ,\label{EQ:WF}
\\
       \mathcal{W}^{-1}(a)   &= 2\int \!\de x\de x^{\prime}\,|x+x^{\prime}\rangle\langle x-x^{\prime}|\int\! \de p ~ a(x,p) e^{\frac{2ipx^{\prime}}{\hbar}}.
\end{align*}
\rem{ 
Conversely, given a phase-space distribution, the Hilbert space operator is recovered via the Weyl transform given by
\begin{equation}
    A  = 2\int \de x\de x^{\prime}\int \de p ~|x+x^{\prime}\rangle a(x,p) e^{2ipx^{\prime}/\hbar} \langle x-x^{\prime}| \,.\label{EQ:WeylTR}
\end{equation}
For notation purposes, one can write equations (\ref{EQ:WF}) and (\ref{EQ:WeylTR}) as
\begin{equation}
a(x,p) = \mathcal{W}(A)
\,, \quad\quad\quad A = \mathcal{W}^{-1}(a) 
\,.
\end{equation}
} 
 \rem{
By introducing the star product \cite{Gro46}\cite{Moyal}, one has the phase-space correspondent of the product of two Hilbert space operators, that is (see equation (112) in \cite{Zachos})
\[
a \star b = 
 \mathcal{W}\left( AB \right)\,.
\]

It is known that the product of Hilbert space operators can be expressed as the Weyl transform of the $\star$-product of the corresponding Wigner distributions via Groenewold's relation \cite{Zachos2005}
\begin{equation*}
AB = \mathcal{W}^{-1}( a\star b )\,,
\end{equation*}
where
\begin{equation*}
 (a\star b)(x,p) = \frac{1}{\hbar^2\pi^2} \int dp^{\prime}dp^{\prime\prime}dx^{\prime}dx^{\prime\prime}~a(x^{\prime},p^{\prime}) b(x^{\prime\prime},p^{\prime\prime}) e^{-\frac{2i}{\hbar}\left[ p(x^{\prime}-x^{\prime\prime}) + p^{\prime}(x^{\prime\prime}-x) + p^{\prime\prime}(x-x^{\prime}) \right]}\,, 
\label{EQ:star}
\end{equation*}
} 
Then, the Moyal bracket (see \cite{moyal1949quantum,Zachos2005} for its explicit definition) is defined in such a way that the commutator between two quantum operators is taken into the Moyal bracket of the corresponding phase space functions, that is \cite{Zachos2005}
\[
\lbrace\!\lbrace a,b \rbrace\!\rbrace = {\frac{1}{i\hbar}  \mathcal{W}\left( [A,B]\right)} 
\,.
\]

At this point, one can express the Lagrangian \eqref{mixedlagr} in terms of phase space functions. Indeed, upon defining the Wigner distribution $W(x,p)=\mathcal{W}(\rho)$ and by replacing the inverse relation $\rho=\mathcal{W}^{-1}(W)$ in \eqref{mixedlagr} one obtains the equivalent  variational principle on phase space
\begin{equation}\label{WMvarprinc}
\delta\int_{t_1}^{t_2}\!\! \iint \de x \de p ~W(x,p) \Big( \hbar \Upsilon(x,p) - H(x,p) \Big)\, \de t=0\,,
\end{equation}
where we have defined  $H(x,p)=\mathcal{W}(H)$ and $\Upsilon(x,p):=\mathcal{W}(i\xi)$.
\rem{
is defined by
\[
    \varUpsilon(x,p) := \frac{1}{\pi\hbar} \int dx^{\prime}~ \langle x+x^{\prime}| i\xi |x-x^{\prime} \rangle e^{-2ipx^{\prime}/\hbar} 
\]
} 
Then, upon recalling the relations \eqref{mixedvariations} and the first in \eqref{variations}, 
one computes
\[
{\delta\Upsilon =\frac{\partial \Theta}{\partial t} + \hbar \lbrace\!\lbrace \Theta, \Upsilon \rbrace\!\rbrace \,,  \quad\quad \delta W = \hbar\lbrace\!\lbrace \Theta, W \rbrace\!\rbrace}\,,  \quad\quad \partial_t W = \hbar\lbrace\!\lbrace \Upsilon, W \rbrace\!\rbrace \,,
\] 
where $\Theta :=\mathcal{W}(i\eta)$, and the Euler-Poincar\'e variational principle \eqref{WMvarprinc} gives
\begin{equation*}
\frac{\partial W(x,p,t)}{\partial t} = \big\lbrace\!\big\lbrace H(x,p), W(x,p,t) \big\rbrace\!\big\rbrace 
\,.
\end{equation*}

Again, we notice that the variational principle \eqref{WMvarprinc} has never appeared before in the literature and it is very different from other approaches proposed earlier, such as \cite{Poulsen11}. In particular, the variational principle \eqref{WMvarprinc} is entirely derived from the Dirac-Frenkel Lagrangian and no  assumption has been made other than unitary evolution.

\subsection{Geodesics on the space of quantum states\label{sec:FS}}
Fubini-Study geodesics on $\boldsymbol{P}\Bbb{C}^n$ are used in various situations of quantum mechanics. Their applications in quantum search algorithms \cite{MiWa01,Zhao2012}, time-optimal control problems \cite{CarHoKoOku06,CarHoKoOku07,Dalessandro01} and holonomic quantum computation \cite{GunWaNa2014,Lucarelli2005,TaNaHa2005} emphasizes their importance and makes this example especially interesting. In the general case, geodesics are optimal curves in the sense that they minimize the distance between two quantum states.

The Fubini-Study geodesics are defined as geodesic equations on $\PH$ minimizing the the Fubini-Study distance. These geodesic flows can be written explicitly as Euler-Lagrange equations associated to the action principle $\delta\int_{t_1}^{t_2}\de t\,L(\psi,\dot\psi)=0$ with Lagrangian $L:T\!\mathscr{H}\to\Bbb{R}$ given by (see e.g. \cite{FaKuMaMa10})
\begin{equation}\label{FSLAGR}
L(\psi,\dot\psi)=\frac{\hbar}{2}\frac{\|\psi\|^2\|\dot\psi\|^2-|\langle\dot\psi|\psi\rangle|^2}{\|\psi\|^4}
\,.
\end{equation}
More explicitly, lengthy computations yield
\[
\big(\|\psi\|^2-\psi\psi^\dagger\big)\big(\|\psi\|^2\ddot\psi-2\langle\psi|\dot\psi\rangle\dot\psi\big)=0
\,.
\]
Notice that this approach does not involve normalized vectors in $\mathscr{H}$. However, a simple way to recover normalization is to use a constrained Lagrangian such as that in \eqref{EQ:constrLagr}.

Application of the Euler-Poincar\'e theory to Fubini-Study geodesics can be performed again by following the procedure outlined in Section \ref{SEC:EPSchr}, without modifications. Then, upon recalling that $\xi^2$ is Hermitian, one obtains the reduced Lagrangian (set $\hbar=1$, for convenience)
\begin{align}
l(\xi,\rho_\psi)
=&-\frac12\left(\left<\rho_\psi,\xi^2\right>+\big<\rho_\psi,i\xi\big>^2\right)
.
\label{FSEPLAGR}
\end{align}
Notice that we recover the well known relation $l(\xi,\rho_\psi)=1/2\,\big(\langle\mathcal{E}^2\rangle- \langle\mathcal{E}\rangle^2\big)$ \cite{AhAn87}, 
where we have introduced the energy operator $\mathcal{E}=i\xi$ and we have used the standard expectation value notation. 
Then, upon  replacing the variational derivatives 
\begin{equation}\label{varderiv}
\frac{\delta l}{\delta\rho_\psi}=
-\frac12\,\xi^2+\big<\rho_\psi|\xi\big>\xi
\,,\qquad\quad
\frac{\delta l}{\delta\xi}=
\frac12\{\rho_\psi,\xi\}-\big<\rho_\psi|\xi\big>\rho_\psi
\end{equation}
in equations \eqref{EQ:EPrho}, standard matrix computations  give
\begin{equation*}
\frac{\de}{\de t}\Big(\{\rho_\psi,\xi\}-2\big<\rho_\psi|\xi\big>\rho_\psi\Big)=0
\,,\qquad\quad
\dot\rho_\psi=[\xi,\rho_\psi]
\,,
\end{equation*}
where the first emphasizes the following conservation form of the Fubini-Study geodesic equation
\[
\frac{\de}{\de t}\Big((\boldsymbol{1}-2\rho_\psi)\dot{\rho}_\psi\Big)=
\frac{\de}{\de t}\!\left(\dot\psi\psi^\dagger-\psi\dot\psi^\dagger-2\langle\psi|\dot\psi\rangle\psi\psi^\dagger\right)=0\,,
\]
as it arises from the left-invariance of the Lagrangian \eqref{FSLAGR} (see e.g. \cite{andersson2013dynamic}). Notice that applying the above conservation law to $\psi$ and  writing $\langle\psi|\dot\psi\rangle=i\langle i\psi,\dot\psi\rangle$ yields
\[
\ddot\psi-\big<\ddot\psi|\psi\big>\psi
=
2i\big<i\psi,\ddot\psi\big>\psi+2\big<\psi|\dot\psi\big>\big(\dot\psi+\big<\dot\psi|\psi\big>\psi\big)
\]
and since $\big<\dot\psi|\psi\big>=-\big<\psi|\dot\psi\big>$, expanding $\big<i\psi,\ddot\psi\big>$ leads to
\[
\big(1-\psi\psi^\dagger\big)\big(\ddot\psi-2\langle\psi|\dot\psi\rangle\dot\psi\big)=0
\,.
\]
This  geodesic flow  can be also recovered as an Euler-Lagrange equation by adding a normalisation constraint $\lambda(\|\psi\|^2-1)$ to the Lagrangian \eqref{FSLAGR} (cf. the constrained DF Lagrangian \eqref{EQ:constrLagr}).

\rem{ 
\begin{remark}[The isoholonomic problem]
Other variants of the Lagrangian \eqref{FSLAGR}  are also studied. For example,  in holonomic quantum computation \cite{}, one inserts the extra constraint $r\langle i\psi,\dot\psi\rangle$ to enforce the condition that the connection form $\langle i\psi,\de\psi\rangle$ vanishes on the Hopf bundle (horizontal lift condition). The so-called \emph{isoholonomic problem} \cite{} is then written as
\[
\delta\int_{t_1}^{t_2\!}\!\left(\frac{\hbar}{2}\frac{\|\psi\|^2\|\dot\psi\|^2-|\langle\dot\psi|\psi\rangle|^2}{\|\psi\|^4}+\lambda(\|\psi\|^2-1)+r\langle i\psi,\dot\psi\rangle\right)\de t=0
\] 
The Euler-Poincar\'e approach applies also in this case without modification, thereby leading to
\begin{equation*}
\frac{\de}{\de t}\Big(\{\rho_\psi,\xi\}+2ir\rho_\psi\Big)=0
\,,\qquad\quad
\dot\rho_\psi=[\xi,\rho_\psi]
\,.
\end{equation*}
Here, the Lagrange multiplier $r(t)$ has to be determined by using the condition $\langle\rho_\psi|\xi\rangle=0$. For example, one may take the trace of the first equation to conclude $\dot{r}=0$. Then, combining the two equations above leads to
\[
\frac{\de}{\de t}\Big((\boldsymbol{1}-2\rho_\psi)\dot{\rho}_\psi+2ir\rho_\psi\Big)=0
\,,\qquad\quad
\langle\rho_\psi|\dot{\rho}_\psi\rangle=0
\,.
\]
\end{remark}
}     

Geodesic flows on the quantum state space have always raised questions concerning their underlying geometric properties \cite{Anandan91,uhlmann2009geometry}. For example, in holonomic quantum computing, a fundamental role is played by the connection form $\langle\psi|\dot\psi\rangle$, whose loop integral defines the celebrated geometric phase. In addition, connection forms also allow the usual horizontal/vertical decomposition on the Hopf bundle. In geometric mechanics, this decomposition can  be performed  by using a more sophisticated theory than Euler-Poincar\'e reduction. This is called \emph{Lagrange-Poincar\'e reduction} \cite{cendra2001lagrangian} and it was recently formulated in the context of homogeneous spaces (arising from symmetry breaking) in \cite{gay2010reduction}. Without entering the technicalities of Lagrange-Poincar\'e reduction, we shall only mention that this theory often takes advantage of a particular connection form that appears to have a precise physical meaning in many different cases: this is called the \emph{mechanical connection} and it is defined in terms of a \emph{momentum map}, another fundamental object in geometric mechanics. This is the topic of the next section.

\subsection{Momentum maps of quantum variational principles\label{sec:momaps}}

The first momentum map one encounters in quantum mechanics is probably the density matrix for pure states \cite{CleMar08}. More precisely, the action of the unitary group $\mathcal{U}(\mathscr{H})$ on the quantum Hilbert space $\mathscr{H}$ (endowed with the symplectic form $\Omega(\psi_1,\psi_2)=2\hbar\left<i\psi_1,\psi_2\right>$) produces the momentum map
\[
J(\psi)=-i\hbar\psi\psi^\dagger\in\mathfrak{u}(\mathscr{H})^*
\,,
\]
as it can be easy obtained by the general formula $\langle J(\psi),\xi\rangle=1/2\,\Omega(\xi\psi,\psi)$ \cite{marsden1999introduction,holm2009geometric}, holding for an arbitrary skew-hermitian operator $\xi\in\mathfrak{u}(\mathscr{H})$. Also, restricting to consider phase transformations yields the total probability or, more precisely, the quantity $J(\psi)=\hbar\|\psi\|^2$.

Other than those above, other momentum map structures appear in geometric quantum dynamics and each correspond to different group actions and different reduction processes. It turns out that in quantum variational principles, the most important momentum map is associated to the action of the isotropy subgroup of the initial state. In order to explain this statement, let us replace the relation \eqref{psievol} in a Lagrangian of the type \eqref{genlagr} and observe that this produces a Lagrangian $\mathcal{L}:T\mathcal{U}(\mathscr{H})\to\Bbb{R}$ by
\begin{equation}\label{callagr-def}
\mathcal{L}(U,\dot{U}):=L(U\psi_0,\dot{U}\psi_0)
\,.
\end{equation}
Although, this Lagrangian is not symmetric under right multiplication, i.e.
\[
\mathcal{L}(U,\dot{U})\neq\mathcal{L}(UU',\dot{U}U')\,,\qquad\quad U'\in\mathcal{U}(\mathscr{H}),
\]
the invariance property is recovered by restricting to the isotropy group of $\psi_0$, that is
\[
\mathcal{U}_{\psi_0}(\mathscr{H})=\{U\in\mathcal{U}(\mathscr{H})\,|\,U\psi_0=\psi_0\}\,.
\]
Indeed, one evidently has
\begin{equation}\label{callagr}
\mathcal{L}(U,\dot{U})=\mathcal{L}(UU_0,\dot{U}U_0)\,,\qquad\quad \forall\,U_0\in\mathcal{U}_{\psi_0}(\mathscr{H}),
\end{equation}
and one may choose the initial vector $\psi_0$ to coincide with the basis vector $\psi_0=(0\,\dots\,0\,1)^\dagger$.
\begin{remark}[Analogy with the heavy top dynamics]
We  observe that the argument above holds in a wide range of situations, including, for example, the Lagrangian reduction for the heavy top dynamics \cite{gay2010reduction}. In that context, the unitary group is replaced by the rotation group $SO(3)$ and the isotropy symmetry is defined to preserve the gravity vector, thereby leading to planar rotations in $SO(2)$. The Noether's conserved quantity (i.e. the momentum map) is then the vertical angular momentum.
\end{remark}

At this point, it is  natural to ask what the momentum map is for the reduction of quantum variational principles. More particularly, we look for the momentum map associated to (the cotangent lift of) the right action of $\mathcal{U}_{\psi_0}(\mathscr{H})$ on the cotangent bundle $T^*\mathcal{U}(\mathscr{H})$. In the general case, it has recently been shown \cite{gay2010reduction} that the momentum map for the right action of a subgroup $G_0\subset G$ on  (the trivialisation of) the cotangent bundle $T^*G\simeq G\times\mathfrak{g}^*$ reads
\begin{equation}\label{genmomap}
J\!\left(g,\mu\right)=\iota^*\!\left(\operatorname{Ad}^*_g\mu\right)
\end{equation}
where $(g,\mu)\in G\times\mathfrak{g}^*$, $\operatorname{Ad}^*_g\mu=g^\dagger\mu g^{-\dagger}$ is the standard matrix coadjoint representation and $\iota^*$ is the dual of the Lie algebra inclusion $\iota:\mathfrak{g}_0\hookrightarrow\mathfrak{g}$. To simplify the treatment, set $\mathscr{H}=\Bbb{C}^n$ and choose $\psi_0=(0\,\dots\,0\,1)^\dagger$ without loss of generality. Then, $\mathcal{U}_{\psi_0}(\mathscr{H})=\mathcal{U}(n-1)\subset\mathcal{U}(n)=\mathcal{U}(\mathscr{H})$ and the group inclusion $\mathcal{U}(n-1)\hookrightarrow\mathcal{U}(n)$ is
\begin{equation}\label{inclusion}
\mathcal{U}(n-1)\ni{U}_0\mapsto\left(\begin{array}{cc}
{U} _0& 0\\
0&1
\end{array}\right)\in \mathcal{U}(n)
\,.
\end{equation}
The corresponding Lie algebra inclusion $\iota:\mathfrak{u}(n-1)\hookrightarrow\mathfrak{u}(n)$ reads
\[
\iota({\xi}_0)=\left(\begin{array}{cc}
{\xi}_0 & 0\\
0&0
\end{array}\right),
\]
while its dual $\iota^*$ is given by $\iota^*(\mu)=(\boldsymbol{1}-\rho_{\psi_0})\mu(\boldsymbol{1}-\rho_{\psi_0})$, that is the standard projection on the upper left block. This result is independent of the number of dimensions and it leads to the following momentum map:
\begin{align}\nonumber
J_1(U,\mu)=&\,\frac12\big\{(\boldsymbol{1}-2\rho_{\psi_0}),\operatorname{Ad}^*_U\mu\big\}+\langle\rho_{\psi_0}|\operatorname{Ad}^*_U\mu\rangle\rho_{\psi_0}
\\
&\,
=\operatorname{Ad}^*_U\!\left(\frac{\delta \ell}{\delta\xi}-\bigg\{\!\rho_{\psi},\frac{\delta \ell}{\delta\xi}\bigg\}+\bigg\langle\rho_{\psi}\bigg|\frac{\delta \ell}{\delta\xi}\bigg\rangle\rho_{\psi}\!\right)
\label{momap1}
\,,
\end{align}
where we have simply rewritten  \eqref{genmomap} by replacing the formula for $\iota^*$. Here, we recall $i\rho_{\psi_0}=\operatorname{Ad}^*_{U}(i\rho_{\psi})$, from the definition $\rho_{\psi}:=\psi\psi^\dagger=U\rho_{\psi_0} U^{-1}$.  Therefore, because of the symmetry property \eqref{callagr} possessed by any Lagrangian of the type \eqref{genlagr}, the corresponding Euler-Poincar\'e equations \eqref{EQ:EPpsi} conserve  $J_{1\!}\!\left(U,{\delta \ell}/{\delta\xi}\right)$. More particularly, one shows that any quantum  system with an arbitrary Lagrangian of the type \eqref{callagr-def} produces dynamics on the zero-level set of $J_{1\!}$. This is easily shown by using the relation $L(\psi,\dot{\psi}) = L(\psi,\xi\psi) =: \ell (\xi,\psi)$, so that 
\begin{equation}\label{mark}
\frac{\delta\ell}{\delta\xi} = \frac{1}{2}\left( \frac{\delta L}{\delta\dot{\psi}}\psi^{\dagger} - \psi\frac{\delta L}{\delta\dot{\psi}}^{\dagger} \right)\,, 
\end{equation}
thereby verifying $J_{1\!}\!\left(U,{\delta \ell}/{\delta\xi}\right)\equiv0$.

Now that we have characterised the momentum map associated to the action of $\mathcal{U}(n-1)$, we recall that all physically relevant Lagrangians must be also phase invariant, so that they can be eventually written in terms of the projection $\rho_\psi=\psi\psi^\dagger\in\mathbf{P}\!\mathscr{H}$. Therefore, the most general symmetry group of the Lagrangian \eqref{callagr-def} has to include phase transformations and this leads us to consider the direct product $\mathcal{U}(n-1)\times\mathcal{U}(1)$. The latter can be embedded in $\mathcal{U}(n)$ by the inclusion
\begin{equation}\label{inclusion2}
\mathcal{U}(n-1)\times\mathcal{U}(1)\ni({U}_0,\varphi)\mapsto\left(\begin{array}{cc}
{U} _0& 0\\
0&e^{-i\varphi}
\end{array}\right)\in \mathcal{U}(n)
\,,
\end{equation}
where the minus sign in the exponent is of purely conventional nature.
Since the momentum map associated to $\mathcal{U}(n-1)$ has already been presented in \eqref{momap1}, we need to compute only the momentum map associated to the group $\mathcal{U}(1)\subset\mathcal{U}(n)$, endowed with the group inclusion
\begin{equation}\label{inclusion3}
\mathcal{U}(1)\ni\varphi\mapsto\left(\begin{array}{cc}
\boldsymbol{1}& 0\\
0&e^{-i\varphi}
\end{array}\right)\in \mathcal{U}(n)
\,.
\end{equation}
Upon computing the dual of the corresponding Lie algebra inclusion $\iota(\alpha)=-i\alpha\rho_{\psi_0}\in\mathfrak{u}(n)$ and by identifying $\mathfrak{u}(1)\simeq\Bbb{R}$, one has the momentum map formula
\begin{equation}\label{momap2}
J_2(U,\mu)=i\left\langle\rho_{\psi_0}\,|\,\operatorname{Ad}^*_U\mu\right\rangle
\,.
\end{equation}
Any quantum  system with an arbitrary Lagrangian of the type \eqref{callagr-def} takes the above momentum map to the form 
\[
J_2(U,\delta\ell/\delta\xi)=\left\langle\psi,i\,\frac{\delta L}{\delta\dot\psi}\right\rangle
=:\mathcal{J}\!\left(\!\psi,\frac{\delta L}{\delta\dot\psi}\right)
,
\]
as it is easily shown by using the relation \eqref{mark}. Here, the momentum map $\mathcal{J}:T^*S(\Bbb{C}^n)\to\Bbb{R}$ arises from the action of $\mathcal{U}(1)$ (notice that we idenitified $\mathfrak{u}(1)\simeq\Bbb{R}$) on the cotangent bundle $T^{*\!}S(\Bbb{C}^n)$. For example, the DF Lagrangian yields $\mathcal{J}(\psi,\delta L/\delta\dot\psi)=\hbar\|\psi\|^2$, while one verifies that the FS Lagrangian \eqref{FSLAGR} makes $\mathcal{J}$ vanish identically.

In more generality, the momentum map corresponding to the action of the full symmetry group $\mathcal{U}(n-1)\times\mathcal{U}(1)$ is given by 
\begin{equation}\label{totalmomap}
J(U,\mu)=\big(J_1(U,\mu), J_2(U,\mu)\big)=\left(\frac12\big\{(\boldsymbol{1}-2\rho_{\psi_0}),\operatorname{Ad}^*_U\mu\big\}+\langle\rho_{\psi_0}|\operatorname{Ad}^*_U\mu\rangle\rho_{\psi_0},\,i\left\langle\rho_{\psi_0}\,|\,\operatorname{Ad}^*_U\mu\right\rangle\right)
\end{equation}
Therefore, because of the symmetry property \eqref{callagr} possessed by any phase-invariant Lagrangian of the type \eqref{genlagr}, the corresponding Euler-Poincar\'e quantum  dynamics \eqref{EQ:EPrho}  conserves the momentum map $J_2$ and lies on the kernel  of $J_1$.

\begin{remark}[Phases and $\mathcal{U}(1)-$actions] Notice that this section has used a $\mathcal{U}(1)-$action that is different from  usual phase transformations. Indeed, while the latter act on vectors by the diagonal action $\psi\mapsto e^{-i\vartheta}\psi$, the $\mathcal{U}(1)-$action used in this section reads $\psi\mapsto (\boldsymbol{1}-\rho_{\psi_0}+e^{-i\varphi}\rho_{\psi_0})\psi$. However, usual phase transformations (denoted by $\mathcal{U}_d(1)$ to emphasise the diagonal action) are a subgroup of $\mathcal{U}(n-1)\times\mathcal{U}(1)$, as it is given by the inclusion
\[
\mathcal{U}_d(1)\ni(\vartheta)\mapsto\left(\begin{array}{cc}
e^{-i\vartheta}\boldsymbol{1}& 0\\
0&e^{-i\vartheta}
\end{array}\right)\in \mathcal{U}(n-1)\times\mathcal{U}(1)\subset\mathcal{U}(n)
\,.
\]
Therefore, our treatment naturally includes the ordinary phase transformations, whose corresponding momentum map is given by $\iota_d^*(J_1(U,\mu),J_2(U,\mu))$. Here, $\iota_d^*$ is the dual of the inclusion $\iota_d:{\mathfrak{u}_d(1)\hookrightarrow\mathfrak{u}(n-1)\times\mathfrak{u}(1)}$ given by $\iota_d(\alpha)=({-i\alpha}\boldsymbol{1},{\alpha})$ (again, we identify $\mathfrak{u}_d(1)\simeq\Bbb{R}$). Upon using the pairing $\langle(\mu,\omega),(\eta,\alpha)\rangle=\langle\mu,\eta\rangle+\omega\alpha$, a direct calculation shows {$\iota_d^*(\mu,\omega)=\omega+\operatorname{Tr}(i\mu)$}, so that $\iota_d^*\big(J_1(U,\mu),J_2(U,\mu)\big)=i\operatorname{Tr}(\mu)$.
\end{remark}

The momentum maps provided in this Section are of paramount importance in geometric quantum dynamics, as they incorporate  essential geometric properties. For example,  it is interesting to notice that the momentum maps \eqref{momap1} and  \eqref{totalmomap} can be used to define connection forms, respectively on the bundles $\mathcal{U}(n)\to S(\Bbb{C}^n)$ and $\mathcal{U}(n)\to \mathbf{P}\Bbb{C}^n$, as they are given in \eqref{bundles}. These connection forms are obtained by applying Lagrange-Poincar\'e reduction for symmetry breaking and we refer the reader to \cite{gay2010reduction} for more details on this topic. In the particular case under consideration, one identifies $\mathfrak{u}(k)^*\simeq\mathfrak{u}(k)$ so that the dual of the inclusion $\iota^*:{\mathfrak{u}(n)^*\to\mathfrak{u}(n-1)^*}$ determines a projection $\Bbb{P}:\mathfrak{u}(n)\to\mathfrak{u}(n-1)$, which in turn can be used to define the \emph{mechanical connection} $\mathcal{A}(\dot{U})={\Bbb{P}(\operatorname{Ad}_{U^{-1}}\xi)}={(\boldsymbol{1}-\rho_{\psi_0})}{(\operatorname{Ad}_{U^{-1}}\xi)}{(\boldsymbol{1}-\rho_{\psi_0})}$ on the bundle  $\mathcal{U}(n)\to S(\Bbb{C}^n)$. An analogous construction yields a connection  on the bundle  $\mathcal{U}(n)\to \mathbf{P}\Bbb{C}^n$, given by $\mathcal{A}(\dot{U})={(\boldsymbol{1}-\rho_{\psi_0})}{(\operatorname{Ad}_{U^{-1}}\xi)}{(\boldsymbol{1}-\rho_{\psi_0})}+i\langle{\operatorname{Ad}_{U^{-1}}\xi\,|\,\rho_{\psi_0}\rangle\rho_{\psi_0}}$. 
Also, it is well known that $\mathcal{A}(\dot\psi)=\langle\psi, i\dot\psi\rangle$ is a principal connection on the Hopf bundle $S(\Bbb{C}^n)\to\mathbf{P}\Bbb{C}^n$. 
Then, the momentum maps presented in this section generate a connection form on each bundle of the diagram below. The study of these connection forms and their curvatures is left for future work.
\begin{diagram}
&  & \mathcal{U}(n) &  &
\\
& \ldTo  &      & \rdTo  &
\\
S(\Bbb{C}^n) &   &   \rTo   &   & \mathbf{P}\Bbb{C}^n
\\
\end{diagram}

The arguments in this Section clarify the meaning of equation \eqref{EQ:projsch}. Indeed, the latter can be interpreted as simply saying that the infinitesimal generator $\xi\in\mathfrak{u}(n)$ (or, equivalently, the  Hamiltonian operator $H$) is defined only up to an element of the isotropy subalgebra $\mathfrak{u}(n-1)\times\mathfrak{u}(1)$. This can be made explicit upon introducing $({\kappa}_0,\alpha_0)\in\mathfrak{u}(n-1)\times\mathfrak{u}(1)$ such that $\{\boldsymbol{1}-2\rho_\psi,\kappa\}={(\boldsymbol{1}-\rho_{\psi})}{\kappa_0}{(\boldsymbol{1}-\rho_{\psi})}+i\alpha_0\rho_\psi$. Although computing $\dot\psi=\xi\psi$ returns the usual phase arbitrariness, as shown in equation \eqref{EQ:projsch2}, the relation \eqref{EQ:projsch} discloses a rich geometry content underlying quantum evolution. Indeed, it reminds that the propagator of quantum dynamics is defined up to elements of $\mathcal{U}(n-1)\times\mathcal{U}(1)$ (not only up to phases in $\mathcal{U}(1)$), which is actually a non Abelian symmetry group (unlike phase transformations). As we shall see in the next Section, this argument accounts for propagators that depend on the initial quantum state $\psi_0$.

\section{The Heisenberg and Dirac pictures}

While the previous sections mainly dealt with the Schr\"odinger picture of quantum mechanics, the Heisenberg picture is rather unexplored in the geometry of quantum evolution. For example, one is interested in the role of projection operators, as they emerge from the projective geometry of the quantum state space in the Schr\"odinger picture. We shall consider the Heisenberg picture for the DF Lagrangian and the Fubini-Study geodesics.

\subsection{Euler-Poincar\'e reduction in the Heisenberg picture\label{sec:HeisDF}}

Most of this Section is devoted to the Heisenberg picture for the DF Lagrangian. It is easy to see that \eqref{DFLAGR} can be written in the Heisenberg picture by introducing
\[
\xi_H:=U^{-1}\dot{U}=\operatorname{Ad}_{U^{-1}} \xi
\,,\qquad\ 
H_H:=U^\dagger H U\,.
\]
Indeed, with these definitions, the  Lagrangian \eqref{DFLAGR} becomes 
\begin{equation}\label{HeisDF}
l(\xi_H,H_H)=\big\langle\rho_{\psi_0}, i\hbar\xi_H-{H}_H\big\rangle\,.
\end{equation}
Then, upon computing
\begin{equation}  \label{variations3} 
\delta\xi_H=\dot\eta_H-[\eta_H,\xi_H]\,,\qquad
\delta H_H=[H_H,\eta_H]
\,,     \qquad\quad 
      \dot{H}_H=[H_H,\xi_H]
\end{equation}
(with $\eta_H:=U^{-1}\delta U$), inserting the Lagrangian \eqref{HeisDF} in the variational principle $\int_{t_1}^{t_2} \!l(\xi_H,H_H)\,\de t=0$ yields the following Euler-Poincar\'e equations:
\begin{equation}\label{generator}
\left[i\hbar\xi_H- H_H,\rho_{\psi_0}\right]=0
\,.
\end{equation}
At this point, we observe that although the above relation is satisfied by $i\hbar\xi_H- H_H=\alpha\boldsymbol{1}$ (so that the Heisenberg Hamiltonian $H_H$ is defined up to a phase factor $\alpha$), more general solutions are present such as 
\begin{equation}\label{generator2}
\xi_H=-i\hbar^{-1\!}H_H+\{\boldsymbol{1}-2\rho_{\psi_0},\kappa\}
\end{equation}
($\kappa$ being arbitrary and skew Hermitian, see equation \eqref{EQ:projsch}). These  solutions have the property of depending on the initial state $\psi_0$. Upon setting $\mathscr{H}=\Bbb{C}^n$ for simplicity, one may choose $\psi_0=(0\,\dots 0\,1)^\dagger$ without loss of generality. Interestingly enough, these more general solutions lead to the unfamiliar equation
\begin{equation}\label{EQ:projsch3}
\dot{H}_H=\left[H_H,\{\boldsymbol{1}-2\rho_{\psi_0},\kappa\}\right],
\end{equation}
so that the Hamiltonian operator $H_H$ is \emph{not} conserved in the general case. Although this may seem surprising, we observe that the above dynamics does not change the physics of the system under consideration. For example,  we observe that the total energy is preserved:
\[
\langle\dot{H}_H\rangle=\langle\rho_{\psi_0}| \dot{H}_H\rangle=0\,,
\]
as shown by a direct verification. Moreover, one realizes that the above dynamics of quantum Hamiltonians returns exactly the \Sch equation \eqref{EQ:projsch2}: indeed, one has
\[
\xi_H\psi_0=-i\hbar^{-1\!}H_H\psi_0-2\langle\rho_{\psi_0}|\kappa\rangle\psi_0
\,
\]
so that,  recalling $\psi=U\psi_0$ and applying $U$ on both sides   returns \eqref{EQ:projsch2}. Notice that it is indeed essential that $\rho_{\psi_0}$  identifies  the initial quantum state. We conclude that the physical content is unaltered by the {Heisenberg equation} \eqref{EQ:projsch3}, which in turn generalizes the standard Heisenberg dynamics (recovered by $\kappa=0$) to incorporate the geometry of quantum dynamics. 

As a practical example, we consider spin dynamics in the Heisenberg picture. In this case, the Hamiltonian reads $H_H=\mathbf{n}\cdot{\boldsymbol{S}}_H$, where ${\boldsymbol{S}}_H(t)=U(t)^{-1}\boldsymbol{S} U(t)$ in standard spin operator notation. The DF Lagrangian \eqref{DFLAGR} is written in the Heisenberg picture as $l(\xi_H,{\boldsymbol{S}}_H)=\big\langle\rho_{\psi_0}, i\hbar\xi_H-\mathbf{n}\cdot{\boldsymbol{S}}_H\big\rangle$, so that the Euler-Poincar\'e equations 
\begin{align*}
\left[i\hbar\xi_H- \mathbf{n}\cdot{\boldsymbol{S}}_H,\,\rho_{\psi_0}\right]=0
\,, 
     \qquad\quad 
      \dot{\boldsymbol\sigma}_H=[{\boldsymbol{S}}_H,\xi_H]
\end{align*}
specialize to yield
\[
\dot{\boldsymbol{S}}_H=i\hbar^{-1\!}\big[\mathbf{n}\cdot{\boldsymbol{S}}_H+i\{2\rho_{\psi_0}-\boldsymbol{1},\kappa\}, {\boldsymbol{S}}_H\big]
=\mathbf{n}\times{\boldsymbol{S}}_H-
\big[\{\boldsymbol{1}-2\rho_{\psi_0},\kappa\}, {\boldsymbol{S}}_H\big]
\]

Notice that this approach can be applied in the general case. For example, one can study linear  oscillator dynamics by recalling the Hamiltonian $H_H=\hbar\omega a^\dagger_H a_H$ and following precisely the same steps as above. The present approach leads to the following Heisenberg equation:
\[
\dot{A}_H=i\hbar^{-1}\big[H_H,A_H\big]-\big[\{\boldsymbol{1}-2\rho_{\psi_0},\kappa\}, A_H\big],
\]
where $\kappa$ is an arbitrary skew-symmetric operator and  $H_H$ undergoes its own evolution \eqref{EQ:projsch3}. In addition, equation \eqref{generator2} yields a new form of the propagator equation
\begin{equation}\label{propagator}
\dot{U}=i\hbar^{-1}HU+U\{\boldsymbol{1}-2\rho_{\psi_0} ,\kappa\}\,.
\end{equation}
It is necessary to point out that, since $\kappa$ is an arbitrary skew-Hermitian matrix parameter, one can simply choose it in such a way that $\{\boldsymbol{1}-2\rho_{\psi_0} ,\kappa\}=0$, thereby eliminating the dependence of the propagator on the initial conditions. A similar argument leads to eliminating phase terms in the Schr\"{o}dinger equation \eqref{EQ:projsch2} \cite{kibble1979geometrization}.

In all this Section, we assumed the initial state is a pure state ${\psi_0}$. If this is not the case, then different solutions of the type $\xi_H=-i\hbar^{-1\!}H_H+\kappa
$ (with $[\kappa,\rho_0]=0$) are allowed by equation \eqref{generator}, because in this case $\rho_{\psi_0}$ is replaced by a density matrix $\rho_0\neq\rho_0^2$.

The Heisenberg picture is particularly natural for the description of FS geodesics. Indeed, while in the \Sch picture the right unitary symmetry is broken by $\rho_\psi$, in the Heisenberg picture one may use the full left symmetry of the Lagrangian \eqref{FSLAGR}, that is $L(\psi,\dot\psi)=L(U^{-1}\psi,U^{-1}\dot\psi)$. Upon recalling \eqref{psievol} and by setting  $\hbar=1$ for convenience, one obtains the Euler-Poincar\'e Lagrangian
\begin{equation}\label{HeisFS}
l(\xi_H)=-{\frac{1}{2}}\Big(\left<\rho_{\psi_0}|\xi_H^2\right>+\left<\rho_{\psi_0},i\xi_H\right>^2\Big)
\,.
\end{equation}
Then, upon using the first of  \eqref{variations2}, the variational principle $\int_{t_1}^{t_2} l(\xi_H)\,\de t=0$ yields
\[
\{\dot{\xi}_H,\rho_{\psi_0}\}-2\big<\rho_{\psi_0}|\dot{\xi}_H\big>\rho_{\psi_0}+\Big[{\xi}_H^2-2\big<\rho_{\psi_0}|\xi_H\big>{\xi}_H,\rho_{\psi_0}\Big]=0\,,
\]
which reflects all the properties already discussed in Section \ref{sec:FS}

\subsection{Dirac-Frenkel Lagrangian in the Dirac picture}
This section extends the arguments from the previous Sections to formulate a new variational principle for quantum dynamics in the Dirac (interaction) picture. As we shall see, the Euler-Poincar\'e construction involves the semidirect product of the unitary group with itself.

In the Dirac picture, the \Sch Hamiltonian operator is split in two parts as $H = H_0 + H_1$, where $H_0$ is typically a simple linear Hamiltonian, while $H_1$ usually contains  nonlinear potential terms. This replacement can then be inserted in the DF Lagrangian \eqref{DFLAGR}. However, it is convenient to keep track of the quantum state $\psi_s$ that is propagated by $H_0$, such that $i\hbar\dot\psi_s=H_0\psi_s$ (up to phase terms). Then, one is led to consider the following DF Lagrangian
\begin{equation}\label{DFLAGR2}
L(\psi,\dot\psi,\psi_s,\dot\psi_s)=\big\langle\psi,i\hbar\dot\psi-({H}_0+H_1)\psi\big\rangle+\big\langle\psi_s,i\hbar\dot\psi_s-{H}_0\psi_s\big\rangle
.
\end{equation} 
Performing Euler-Poincar\'e reduction by replacing the evolution relation \eqref{psievol} on the first part yields the Lagrangian
\[
\bar{L}(\xi,\rho_\psi,\psi_s,\dot\psi_s)=\left\langle\rho_\psi,i\hbar\xi-{H}_0-H_1\right\rangle+\big\langle\psi_s,i\hbar\dot\psi_s-{H}_0\psi_s\big\rangle
\,,
\]
with $\rho_\psi=U\rho_{\psi_0}U^{-1}$ and $\xi:=\dot{U}U^{-1}$.
At this point, the propagator associated to $H_0$ can be used to replace the evolution relation $\psi_s(t)=U_0(t)\bar{\psi}_0$ in the second term, thereby leading to the Euler-Poincar\'e Lagrangian
\begin{equation}\label{DFLagrInt}
l(\xi_0,\xi_I,\rho_{\psi_I},H_{0,I},H_{1,I})=\left\langle\rho_{\psi_I},i\hbar\xi_I-H_{0,I}-H_{1,I}\right\rangle+\big\langle\rho_{\bar\psi_0},i\hbar\xi_0-{H}_{0,I}\big\rangle
\end{equation}
where $\rho_{\bar\psi_0}=\bar{\psi}_0\bar{\psi}_0^\dagger$  and we have introduced the following definitions
\begin{equation}\label{defs}
\psi_I=U_0^{-1}\psi
\,,\qquad
H_{j,I}=U_0^{-1}H_{j}U_0
\,,\qquad
\xi_0={U}_0^{-1}\dot{U}_0
\,,\qquad
\xi_I=U_0^{-1}\xi U_0=\operatorname{Ad}_{U_0^{-1}\!}\xi
\,.
\end{equation}

In order to write the resulting equations of motion, we start by using the last two definitions in \eqref{defs} to compute the variations
\begin{equation}\label{semidvars}
\delta(\xi_0,\xi_I)=\big(\dot{\eta}_0+[\xi_0,\eta_0], \dot{\eta}_I+[\xi_0,\eta_I]-[\eta_0,\xi_I]+[\eta_I,\xi_I]\big)
\end{equation}
where $\eta_0=U_0^{-1}\delta U_0$ and $\eta_I=\operatorname{Ad}_{U_0^{-1}}((\delta U) U^{-1})$. One recognizes that the variations \eqref{semidvars} are Euler-Poincar\'e variations of the type $\delta\nu=\dot{\zeta}+[\zeta,\nu]_\mathfrak{g}$, where $[\cdot\,,\cdot]_\mathfrak{g}$ is the Lie bracket on $\mathfrak{g}=\mathfrak{u}_{0}(\mathscr{H})\,\circledS\,\mathfrak{u}(\mathscr{H})$, that is the Lie algebra of the  semidirect product group $\mathcal{U}_{0}(\mathscr{H})\,\circledS\,\mathcal{U}(\mathscr{H})$. Here, the group $\mathcal{U}_{0}(\mathscr{H})$ is a copy of the unitary group $\mathcal{U}(\mathscr{H})$ and one thinks of $\mathcal{U}_{0}(\mathscr{H})$ as accounting for the propagators $U_0\in\mathcal{U}_{0}(\mathscr{H})$.

Computation of the other variations by using \eqref{defs} yields
\[
\delta\rho_{\psi_I}=[\eta_I-\eta_0,\rho_{\psi_I}]
\,,\qquad
\delta H_{j,I}=[H_{j,I},\eta_0]
\,,
\]
so that the variational principle $\int_{t_1}^{t_2\!}l(\xi_0,\xi_I,\rho_I,H_{0,I},H_{1,I})\,\de t=0$ produces the following equations of motion for an arbitrary Lagrangian $l$:
\begin{align}
    &\frac{\de}{\de t}\frac{\delta l}{\delta\xi_{I}} - \left[ \xi_{I},\frac{\delta l}{\delta\xi_{I}} \right] + \left[ \xi_{0},\frac{\delta l}{\delta \xi_{I}}\right] + \left[\rho_{\psi_I},\frac{\delta l}{\delta\rho_{\psi_I}} \right]
    = 0\,, \label{EQ:DiracEP1} 
    \\
    &\frac{\de}{\de t}\frac{\delta l}{\delta\xi_{0}} + \left[ \xi_{0},\frac{\delta l}{\delta\xi_{0}} \right] + \left[ \xi_{I},\frac{\delta l}{\delta\xi_{I}} \right] 
    -\left[\rho_{\psi_I},\frac{\delta l}{\delta\rho_{\psi_I}} \right]
    - \left[ H_{0,I},\frac{\delta l}{\delta H_{0,I}} \right] - \left[ H_{1,I},\frac{\delta l}{\delta H_{1,I}} \right] = 0 \,,
    \label{EQ:DiracEP2}
    \\
    &
\dot{\rho}_{I} = [\xi_{I}- \xi_{0},\rho_{I}] \,, \qquad
\dot{H}_{j,I} = [H_{j,I},\xi_{0}]\,.
\label{EQ:DiracEP3}
\end{align}
Then, computing the variational derivatives of the Lagrangian \eqref{DFLagrInt} and replacing them into \eqref{EQ:DiracEP1} and \eqref{EQ:DiracEP2} gives
\[
i\hbar\big[ \xi_{I}, \rho_{\psi_I} \big]= \big[H_{0,I}+H_{1,I}, \rho_{\psi_I} \big] 
\,,\qquad\ 
\big[ i\hbar\xi_{0}-H_{0,I},\rho_{\bar\psi_0} \big]=0
\]
As seen in Section \ref{sec:HeisDF}, the second relation above is solved by
\[
\xi_0=-i\hbar^{-1\!}H_{0,I}+\{\boldsymbol{1}-2\rho_{\bar\psi_0},\kappa\}
\,,
\]
so the second in  \eqref{EQ:DiracEP3} gives
\[
\dot{H}_{0,I} = [H_{0,I},\{\boldsymbol{1}-2\rho_{\bar\psi_0},\kappa\}]
\,,\qquad
\dot{H}_{1,I} = i\hbar^{-1\!}[H_{0,I},H_{1,I}]+[H_{1,I},\{\boldsymbol{1}-2\rho_{\bar\psi_0},\kappa\}]\,.
\]
On the other hand, the first in \eqref{EQ:DiracEP3} becomes
\[
\dot{\rho}_{\psi_I}=i\hbar^{-1}\big[{\rho}_{\psi_I},H_{1,I}\big]+\big[{\rho}_{\psi_I},\{\boldsymbol{1}-2\rho_{\bar\psi_0},\kappa\}\big]
\,.
\]
Then, we notice that the choice $\kappa=0$ returns the usual quantum dynamics in the Dirac picture. As we know from Section \ref{sec:HeisDF}, the $\kappa$-terms do not change the overall physical content of the dynamics. For example, a direct calculation verifies the following energy conservations:
\[
\frac{\de}{\de t}\,\big\langle\rho_{\psi_I}|H_{0,I}+H_{1,I}\big\rangle=0
\,,\qquad\ 
\frac{\de}{\de t}\,\big\langle\rho_{\bar\psi_0}|H_{0,I}\big\rangle=0
\,.
\]

As we have seen, the application of Euler-Poincar\'e reduction theory reveals the geometric features emerging in the Heisenberg picture of quantum dynamics. These geometric features reveal the  form \eqref{propagator} of the propagator equation, without affecting the physical content of quantum dynamics. When this  form of the propagator equation is considered in the Dirac picture, this introduces extra terms in the dynamics, which still preserve the total energy of the system. 

In the next Section, we shall formulate a new variational principle for the interaction of classical and quantum degrees of freedom.

\section{Classical-quantum variational principles}

The interplay of quantum and classical degrees of freedom has always attracted much attention in quantum mechanics. For example, the consistent formulation of hybrid quantum-classical models in molecular dynamics remains an outstanding issue \cite{Salcedo2012}. In this section, we present a geometric formulation of the most elementary system coupling classical and quantum dynamics. This is given by combining the Ehrenfest equations for the expectation of the canonical variables with the Schr\"odinger/Liouville equation for the quantum degrees of freedom. More particularly, we shall present a novel variational principle for the Ehrenfest mean field model and in more generality for expectation value dynamics.

\subsection{The classical-quantum mean field model}
In order to approach the dynamics of quantum expectations, we observe that the mean field closure of any classical-quantum system can be derived in first instance by the following Lagrangian
\begin{equation}\label{miguel}
L(\bz,\dot\bz,\psi,\dot\psi)=\frac12\,\dot\bz\cdot\Bbb{J}\bz+\langle\psi, i\hbar\dot\psi - H(\bz)\psi\rangle
\,,
\end{equation}
where $-\Bbb{J}_{ij\,}\de z^i\wedge\de z^j$ is the canonical symplectic form and $H(\bz)$ is a Hermitian operator depending on the classical degrees of freedom $\bz=(\bx,\bp)$. Indeed, the corresponding Euler-Lagrange equations yield
\begin{equation}
\dot\bz=\Bbb{J}\left\langle\psi|\nabla_\bz H(\bz)\psi\right\rangle
\,,\qquad\qquad
i\hbar\dot\psi=H(\bz)\psi\,,
\label{HybridEqs1}
\end{equation}
thereby recovering the ordinary mean field model of classical-quantum dynamics (see e.g. equations (12.2)-(12.4) in \cite{Wyatt2006}). Here, purely classical dynamics is recovered by the phase type Hamiltonian $H(\bz)={\sf h}(\bz)I$ (here, $I$ denotes the identity operator on $\mathscr{H}$), while purely quantum dynamics is recovered when ${\nabla_\bz H(\bz)=0}$.

An Euler-Poincar\'e formulation of the above equations can again be obtained by letting the quantum state evolve under unitary transformations. This leads to a coupled Euler-Lagrange equation for $(\bz,\dot\bz)$ and the Euler-Poincar\'e equations for the quantum dynamics (expressed in terms of either $\psi$ or its density matrix). However, in order to find a full set of Euler-Poincar\'e equations that includes the classical evolution, we may choose to evolve the phase-space vector $\bz$ under the action of the Heisenberg group (i.e., phase-space translations), which is prominent in the theory of quantum coherent states. To this purpose, consider a curve $h(t)=(\mathbf{h}(t),\varphi(t))$ in the Heisenberg group $\mathcal{H}(\Bbb{R}^{2n})\simeq \Bbb{R}^{2n+1}$ and let the phase space vector $\bz$ evolve as
\begin{equation}\label{DirectProdAct}
\bz(t)=\bz_0+\mathbf{h}(t)
\,.
\end{equation}
Also, we recall $\psi(t)=U(t)\psi_0$. Then, upon inserting the auxiliary phase factor $\varphi$ in the Lagrangian \eqref{miguel}, the latter becomes
\[
L_{\bz_0,\psi_0}({h},\dot{{h}},U,\dot{U})=\frac12\,\dot{\mathbf{h}}\cdot\Bbb{J}(\bz_0+\mathbf{h})+\big\langle U\psi_0, \big(i\hbar\dot{U}+\dot\varphi U - H(\bz_0+\mathbf{h})U\big)\psi_0\big\rangle
\,.
\]
The above Lagrangian is of the type 
\[
L_{\bz_0,\psi_0}:T\mathcal{H}(\Bbb{R}^{2n})\!\times T\mathcal{U}(\mathscr{H})\to\Bbb{R}
\]
and its dynamics can be approached by Euler-Poincar\'e reduction. Therefore, in order to find an expression for the reduced variable $\zeta:=\dot{h}h^{-1}$, we define the Lie algebra element
\[
(\bzeta,\phi):=\left.\frac{\de}{\de s}\right|_{s=0\!}\left(g(s)h^{-1}\right)
\in\mathfrak{h}(\Bbb{R}^{2n})
\]
where $g(s)=(\mathbf{g}(s),\vartheta(s))\in\mathcal{H}(\Bbb{R}^{2n})$ is a curve such that $g(0)=h$ and ${g}'(0)=\dot{h}$ (for some fixed time). Here, we  recall the Heisenberg group operation
\begin{equation}\label{Heismult}
gh
=\left(\mathbf{g}+\mathbf{h},\, \vartheta+\varphi+\frac12\,\mathbf{g}\cdot\Bbb{J}\mathbf{h}\right)
\,,\qquad
\qquad
\forall
g,h\in\mathcal{H}(\Bbb{R}^{2n})
\,, 
\end{equation}
which gives $h^{-1}=(-\mathbf{h},-\varphi)$. Eventually, one finds
\[
\zeta=(\bzeta,\phi)=\left(\dot{\mathbf{h}},\,\dot{\varphi}-\frac12\,\dot{\mathbf{h}}\cdot\Bbb{J}\mathbf{h}\right)
,
\]
so that the Euler-Poincar\'e Lagrangian is written as
\begin{equation}
\label{HybridLagrangian1}
\ell(\boldsymbol\zeta,\phi,\xi,\bz,\rho_\psi)=\bzeta\cdot\Bbb{J}\bz +\big\langle\rho_\psi,i\hbar\xi+\phi-H(\bz)\big\rangle
\,,
\end{equation}
where we have used the convenient initial condition $\bz_0=0$ in \eqref{DirectProdAct}, without loss of generality. 
Notice, this Lagrangian is of the type
\[
\ell:\big(\mathfrak{h}(\Bbb{R}^{2n})\oplus\mathfrak{u}(\mathscr{H})\big)\times\big(\Bbb{R}^{2n}\times\mathbf{P}_{\!}\mathscr{H}\big)\to\Bbb{R}
\,.
\]
Then, the Euler-Poincar\'e equations follow in the theorem below, upon recalling the infinitesimal adjoint representation 
\begin{equation}\label{InfAct}
\operatorname{ad}_{(\boldsymbol\zeta_1,\phi_1)}(\boldsymbol\zeta_2,\phi_2)=(0,- \bzeta_1\cdot{\Bbb{J}\bzeta_2})
\end{equation}
in the Heisenberg Lie algebra $\mathfrak{h}(\Bbb{R}^{2n})$.
The present treatment is now extended to the case of mixed quantum states.
\begin{theorem}
Consider the variational principle
\[
\delta\int_{t_1}^{t_2}\!\Big(\bzeta\cdot\Bbb{J}\bz +\big\langle\rho,i\hbar\xi+\phi-H(\bz)\big\rangle\Big)\,\de t = 0
\]
 and the variations
\[
(\delta\boldsymbol\zeta,\delta\phi) = 
\big(\dot{\boldsymbol\gamma},\,\dot{\theta}+\boldsymbol\zeta\cdot\Bbb{J}\boldsymbol\gamma\big)
\,,\qquad
\delta\xi=\dot{\eta}-[\xi,\eta]
\,
\,, 
\qquad \delta\bz = \boldsymbol{\gamma}\,,
\qquad
\delta\rho = [\eta,\rho]
\,,
\]
where $({\boldsymbol\gamma},{\theta})$ and $\eta$ are arbitrary and vanish at the endpoints. Together with the auxiliary equations
\[
\dot\bz=\bzeta\,,\qquad
\dot\rho=[\xi,\rho]
\,,
\]
this variational principle is equivalent to the equations of motion
\begin{equation*}
\dot{\bz} = \Bbb{J}\langle \rho | \nabla_{\bz}H(\bz) \rangle 
\,,\qquad\qquad i\hbar\dot{\rho} = \left[ H(\bz),\rho \right]\,.
\end{equation*}
\end{theorem}
\paragraph{Proof.}
Consider the general Lagrangian of the form $\ell(\boldsymbol\zeta,\phi,\xi,\bz,\rho_\psi)$. By direct substitution of the variations into the variational principle 
\begin{align*}
\delta\int_{t_1}^{t_2}\!\Big( \left\langle \frac{\delta\ell}{\delta\bzeta}, \, \dot{\boldsymbol\gamma} \right\rangle 
+ \left\langle \frac{\delta\ell}{\delta\phi}, \, \dot{\theta}+\boldsymbol\zeta\cdot\Bbb{J}\boldsymbol\gamma \right\rangle   
+ \left\langle \frac{\delta\ell}{\delta\xi}, \, \dot{\eta}-[\xi,\eta] \right\rangle
+ \left\langle \frac{\delta\ell}{\delta\bz}, \,  \boldsymbol{\gamma} \right\rangle
+ \left\langle \frac{\delta\ell}{\delta\rho}, \,  [\eta,\rho] \right\rangle
\Big) \de t = 0\,,
\end{align*}
one writes the Euler-Poincar\'e equations as
\[
-\frac{\de}{\de t}\frac{\delta\ell}{\delta\bzeta} + \frac{\delta\ell}{\delta\phi}\Bbb{J}\bzeta + \frac{\delta\ell}{\delta\bz} = 0\,, \qquad 
\frac{\de}{\de t} \frac{\delta\ell}{\delta\phi} = 0\,, \qquad 
-\frac{\de}{\de t} \frac{\delta\ell}{\delta\xi} + \left[ \xi,\frac{\delta\ell}{\delta\xi} \right] + \left[ \rho, \frac{\delta\ell}{\delta\rho} \right] =0\,.
\]
In particular, for the Lagrangian \eqref{HybridLagrangian1}, we have
\[
\frac{\delta\ell}{\delta\bzeta} = \Bbb{J}\bz\,, \qquad 
\frac{\delta\ell}{\delta\phi} = \langle \rho |1 \rangle\,, \qquad
\frac{\delta\ell}{\delta\xi} = -i\hbar \rho\,, \qquad
\frac{\delta\ell}{\delta\rho} = i\hbar\xi + \phi - H(\bz)\,, \qquad
\frac{\delta\ell}{\delta\bz} = -\Bbb{J}\bzeta - \langle \rho | \nabla_{\bz}H(\bz) \rangle\,,
\]
such that the Euler-Poincar\'e equations yield
\begin{align*}
-\Bbb{J}\dot{\bz} + \Bbb{J}\bzeta - \Bbb{J}\bzeta - \langle \rho | \nabla_{\bz}H(\bz) \rangle& = 0  
\\
-i\hbar\dot{\rho} + \left[ \xi,-i\hbar\rho \right] - \left[ \rho, i\hbar\xi - H(\bz) \right] &=0  \,.
\end{align*}
thereby completeing the proof. \quad$\blacksquare$

\noindent
The observation that hybrid classical-quantum dynamics can be expressed by using the Heisenberg and unitary groups motivates us to investigate further the interplay between these two symmetry structures. The next Section shows that combining the two groups into a semidirect product yields the variational formulation of quantum expectation dynamics.

\subsection{The semidirect product $\mathcal{H}(\Bbb{R}^{2n})\,\circledS\; \mathcal{U}(\mathscr{H})$}

While the previous Section used the direct product $\mathcal{H}(\Bbb{R}^{2n})\!\times\mathcal{U}(\mathscr{H})$ group structure to obtain hybrid classical-quantum dynamics, we shall now illustrate how constructing the semidirect product $\mathcal{H}(\Bbb{R}^{2n})\,\circledS\,\mathcal{U}(\mathscr{H})$ allows to shed new light on the dynamics of expectation values, thereby extending Ehrenfest theorem to more general situations.

The semidirect product $\mathcal{H}(\Bbb{R}^{2n})\,\circledS\,\mathcal{U}(\mathscr{H})$ can be constructed upon using the celebrated displacement operator from the theory of coherent quantum states. This is defined as follows
\[
U_h\psi(\boldsymbol{x})=e^{-\frac{i}{\hbar}\big(\varphi+\frac{\mathbf{h}_p\cdot\mathbf{h}_q}{2}-\mathbf{h}_p\cdot\boldsymbol{x}\big)}\psi(\boldsymbol{x}-\mathbf{h}_q)
\,,\qquad
\qquad
\forall
h=(\mathbf{h},\varphi)\in\mathcal{H}(\Bbb{R}^{2n})
\,,
\]
where the phase space vector $\mathbf{h}\in\Bbb{R}^{2n}$ is expressed as $\mathbf{h}=(\mathbf{h}_q,\mathbf{h}_p)$. This operator defines a group homomorphism that can be used to construct the following product rule in $\mathcal{H}(\Bbb{R}^{2n})\,\circledS\,\mathcal{U}(\mathscr{H})$:
\begin{equation} \label{productrule}
(h_1,U_1)(h_2,U_2)=\big(h_1h_2,U_1({U_{h_1}}U_2U_{h_1}^\dagger)\big)
\,,\qquad
\qquad
\forall
h_1,h_2\in\mathcal{H}(\Bbb{R}^{2n})
\,,
\quad
\forall U_1,U_2\in\mathcal{U}(\mathscr{H})
\,,
\end{equation} 
where $h_1h_2$ is the product rule in the Heisenberg group, already defined in \eqref{Heismult}. Notice, upon denoting ${Z}=({Q},{P})$ (quantum canonical operators), the displacement operator $U_h$ leads to the following Lie algebra homomorphism $\iota:\mathfrak{h}(\Bbb{R}^{2n})\to\mathfrak{u}(\mathscr{H})$
\[
\iota(\zeta)=-i\hbar^{-1}(\phi+\boldsymbol{\zeta}\cdot\Bbb{J}{Z})
\,,\qquad
\qquad
\forall
\zeta=(\boldsymbol{\zeta},\phi)\in\mathfrak{h}(\Bbb{R}^{2n})\,,
\]
which occurs in the Lie bracket structure on $\mathfrak{h}(\Bbb{R}^{2n})\,\circledS\,\mathfrak{u}(\mathscr{H})$, given by
\[
\operatorname{ad}_{(\zeta_1,\xi_1)}(\zeta_2,\xi_2)=\Big(\operatorname{ad}_{\zeta_1\!}\zeta_2,\, [\xi_1,\iota(\zeta_2)]-[\xi_2,\iota(\zeta_1)]+[\xi_1,\xi_2]\Big)
\,,
\]
where the operator `$\operatorname{ad}$' appearing in the first slot on the RHS is the infinitesimal adjoint action on $\mathfrak{h}(\Bbb{R}^{2n})$, as it was defined in \eqref{InfAct}. No confusion should arise from this notation.

In order to construct a dynamical theory by using the group structure above in the Lagrangian \eqref{miguel}, we need to find an action of $\mathcal{H}(\Bbb{R}^{2n})\,\circledS\,\mathcal{U}(\mathscr{H})$ on the space $\Bbb{R}^{2n\!}\times S(\mathscr{H})$. This task can be achieved by computing the coadjoint representation on the semidirect product. This computation can benefit from the following property.
\begin{lemma}[Equivariance] \label{equivariance}
With the notation above, the following relations hold
\[
U_h Z U_h^\dagger= Z-\mathbf{h}{I}
\,,
\qquad\qquad
\iota(\operatorname{Ad}_{h}\zeta)={U_h\,}\iota(\zeta)\,U_h^\dagger\,,
\]
where ${I}$ is the identity operator on $\mathscr{H}$ and  $\operatorname{Ad}_h\zeta=\left( \boldsymbol{\zeta}, \,\phi + \mathbf{h} \cdot\Bbb{J}\boldsymbol{\zeta} \right)$ is the adjoint representation on $\mathcal{H}(\Bbb{R}^{2n})$.
\end{lemma}
\paragraph{Proof.}
The first relation is easily proved by a direct verification. The first component reads as follows:
\begin{align*}
\big(U_h{X}U_h^\dagger\big)\psi(\boldsymbol{x}) &= \big( U_h \boldsymbol{x}\big)\left[ e^{\frac{i}{\hbar}\varphi} e^{-i\frac{\mathbf{h}_p\cdot\mathbf{h}_q}{2\hbar}}  e^{-i\frac{\mathbf{h}_p\cdot\boldsymbol{x}}{\hbar}} \psi(\boldsymbol{x}+\mathbf{h}_q) \right]
\\
&=e^{-i\frac{\mathbf{h}_p\cdot\mathbf{h}_q}{\hbar}}  e^{i\frac{\mathbf{h}_p\cdot\boldsymbol{x}}{\hbar}}  ( \boldsymbol{x}-\mathbf{h}_q ) e^{-i\frac{\mathbf{h}_p\cdot (\boldsymbol{x}-\mathbf{h}_q)}{\hbar}} \psi(\boldsymbol{x})
\\
&=  ( \boldsymbol{x}-\mathbf{h}_q )  \psi(\boldsymbol{x})
\,.
\end{align*}
Similarly, the second component reads
\begin{align*}
\big(U_h{P}U_h^\dagger\big)\psi(\boldsymbol{x}) &= -i\hbar\,\big(  U_{h} \nabla \big)\left[ e^{\frac{i}{\hbar}\varphi} e^{-i\frac{\mathbf{h}_p\cdot\mathbf{h}_q}{2\hbar}}  e^{-i\frac{\mathbf{h}_p\cdot\boldsymbol{x}}{\hbar}} \psi(\boldsymbol{x}+\mathbf{h}_q) \right]
\\
&= -  U_h  \Big[ e^{\frac{i}{\hbar}\varphi} e^{-i\frac{\mathbf{h}_p\cdot\mathbf{h}_q}{2\hbar}} e^{-i\frac{\mathbf{h}_p\cdot\boldsymbol{x}}{\hbar}} \big({\mathbf{h}_p}\, \psi(\boldsymbol{x}+\mathbf{h}_q) + i\hbar \nabla\psi(\boldsymbol{x}+\mathbf{h}_q)  \big)\Big]
\\
&=
(-{\mathbf{h}_p}+P)  \psi(\boldsymbol{x})
\,.
\end{align*}
Combining both components proves the first relation in the lemma.
The second relation follows by direct substitution
\begin{multline*}
\iota(\operatorname{Ad}_{h}\zeta) = \iota\Big( \boldsymbol{\zeta}, \,\phi + \mathbf{h} \cdot\Bbb{J}\boldsymbol{\zeta} \Big) = -i\hbar^{-1}\Big( \phi + \mathbf{h}\cdot\Bbb{J}\boldsymbol{\zeta} - Z\cdot\Bbb{J}\boldsymbol{\zeta} \Big) = -i\hbar \left( \phi - \left( Z - \mathbf{h}{I} \right)\cdot \Bbb{J}\bzeta \right) 
=  
\\
 = -i\hbar \big( \phi - \big( U_h Z U_h^\dagger\big)\cdot \Bbb{J}\bzeta \big) 
=  U_h\big(-i\hbar \big( \phi +\Bbb{J} Z \cdot \bzeta \big) \big) U_h^\dagger 
={U_h\,}\iota(\zeta)\,U_h^\dagger
\,,
\end{multline*}
thereby completing the proof.
$\quad\blacksquare$

\noindent
Eventually, by making use of the previous relations in the definition of coadjoint representation, one finds the following expression:
\[
\operatorname{Ad}^*_{(h,U)}(\nu,\mu)=\Big(\boldsymbol\nu-\alpha\Bbb{J}\mathbf{h}+\big\langle\mu-U^\dagger\mu U,i\hbar^{-1}\Bbb{J}Z\big\rangle, \, \alpha,\,U_h^\dagger U^\dagger\mu UU_h\Big)
\]
where we have used the notation $\nu=(\boldsymbol\nu,\alpha)\in \mathfrak{h}(\Bbb{R}^{2n})^*\simeq\Bbb{R}^{2n+1}$. This coadjoint representation is computed explicilty in the Appendix  \ref{Appendix1}.

Then, upon fixing the invariant set $\alpha=1$ and by introducing the variables $\bz=-\Bbb{J}\boldsymbol\nu$ and $\rho_\psi=i\hbar^{-1}\mu$, we obtain the following action of $\mathcal{H}(\Bbb{R}^{2n})\,\circledS\,\mathcal{U}(\mathscr{H})$ on the space $\Bbb{R}^{2n\!}\times \mathbf{P}_{\!}\mathscr{H}$:
\[
\Phi_{(h,U)}(\bz,\rho_\psi)=\Big(\bz-\mathbf{h}+\langle UZU^\dagger-Z \rangle,\,U_h^\dagger  U^\dagger\rho_\psi U U_h\Big)
\,,
\]
where we have used the expectation value notation $\langle A\rangle=\langle A | \rho_\psi\rangle$.

\subsection{Geometry of quantum expectation dynamics}

At this point, the semidirect product $\mathcal{H}(\Bbb{R}^{2n})\,\circledS\; \mathcal{U}(\mathscr{H})$ has been characterized and it has been showed to possess an action on the classical-quantum phase space $\Bbb{R}^{2n\!}\times \mathbf{P}_{\!}\mathscr{H}$. Then, we consider the evolution of the classical-quantum variables $(\bz,\rho_\psi)$ under the action of $(h^{-1},U^{-1})$, which then gives
\begin{equation}\label{SemiDEvol}
\bz(t)=\bz_0+\mathbf{h}(t)+\big\langle U(t)^\dagger ZU(t)-Z\,\big|\,\rho_{\psi_0} \big\rangle
\,,\qquad\quad
\rho_\psi(t)=U_h(t) U(t)\rho_{\psi_0}U(t)^\dagger U_h(t)^\dagger
\,.
\end{equation}
The evolution above  has the following crucial feature:
\[
\bz(t) -\big\langle Z\big|\rho_{\psi}(t) \big\rangle=\bz_0
 -\langle Z|\rho_{\psi_0} \rangle\,,
\]
as it is verified upon computing
\[
\big\langle  Z\,\big|\,U(t)\rho_{\psi_0}U(t)^\dagger \big\rangle=
\big\langle  Z\,\big|\,U_h(t)^\dagger\rho_{\psi}(t)U_h(t) \big\rangle
=
\big\langle  U_h(t) ZU_h(t)^\dagger\,\big|\,\rho_{\psi}(t) \big\rangle=
\big\langle  Z-\mathbf{h}(t)I\,\big|\,\rho_{\psi}(t) \big\rangle\,.
\]
Therefore, in order to study expectation value dynamics, one can simply initiate the evolution under the initial condition $\bz_0=\langle Z\,|\,\rho_{\psi_0} \rangle$, which is then replaced in \eqref{SemiDEvol}. Moreover, the evolution above, produces the equations of motion
\[
\dot{\bz} = \boldsymbol{\zeta} - \big\langle [\rho,{Z}],\xi \big\rangle\,, 
\qquad\quad
\dot\rho =  \left[ i\hbar^{-1}{Z}\cdot\Bbb{J}\boldsymbol{\zeta}+ \xi,\rho \right]
\]
where $\zeta=\dot{\mathbf{h}}$ and $\xi=U_{h}\dot{U}U^{\dagger}U_{h}^{\dagger}$.  Analogous expressions hold for the variations $(\delta\bz,\delta\rho)$.

At this point, we consider the Euler-Poincar\'e Lagrangian of the classical-quantum mean field model \eqref{HybridLagrangian1}. Although that was written previously on the space $(\mathfrak{h}(\Bbb{R}^{2n})\oplus\mathfrak{u}(\mathscr{H}))\times(\Bbb{R}^{2n}\times\mathbf{P}_{\!}\mathscr{H})$, we now change perspective and we interpret the same expression \eqref{HybridLagrangian1} for $\ell(\boldsymbol\zeta,\phi,\xi,\bz,\rho_\psi)$ as a Lagrangian of the type
\[
\ell:\big(\mathfrak{h}(\Bbb{R}^{2n})\,\circledS\,\mathfrak{u}(\mathscr{H})\big)\times\big(\Bbb{R}^{2n}\times\mathbf{P}_{\!}\mathscr{H}\big)\to\Bbb{R}
\,.
\]
Notice that the Hamiltonian operator $H(\bz)$ depends on the classical variable $\bz$, which has to be interpreted as the expectation value $\langle Z|\rho_{\psi}\rangle$. This amounts to consider quantum systems for which the total energy can be written in terms of both the quantum state $\rho_\psi$ and its corresponding expectation values $\bz=\langle Z|\rho_{\psi}\rangle$. (Notice that this is a very general case, as it shown by considering the kinetic energy expression $\langle P^2\rangle_\psi/2=\langle p\rangle^2/2+\langle P-\langle p\rangle\rangle_{\!\psi}^{\,2}/2$).

\begin{theorem} 
Consider the Lagrangian  \eqref{HybridLagrangian1} and its associated variational principle for mixed quantum states
\[
\delta\int_{t_1}^{t_2}\bigg(\bzeta(t)\cdot\Bbb{J}\bz(t) +\Big\langle\rho(t),i\hbar\xi(t)+\phi(t)-H(\bz(t))\Big\rangle\bigg)\de t=0
\,,
\]
with variations
\begin{align*}
\delta\bzeta &= \dot{\boldsymbol{\gamma}}\,, 
\qquad\  \delta\phi = \dot{\theta} - \bzeta\cdot\Bbb{J}\boldsymbol{\gamma}\,,
\\
\delta\xi &= \dot{\eta} - i\hbar^{-1}\big(\left[\xi,{Z}\cdot\Bbb{J}\boldsymbol{\gamma}\right] - \left[ \eta,{Z}\cdot\Bbb{J}\boldsymbol{\zeta} \right]\big) + [\eta,{\xi}]\,,
\\
\delta\bz &= \boldsymbol{\gamma} - \big\langle [\rho,Z],\eta \big\rangle\,, 
\\
\delta\rho &=  \left[ i\hbar^{-1}{Z}\cdot\Bbb{J}\boldsymbol{\gamma}+ \eta,\rho \right]\,,
\end{align*}
where $\boldsymbol{\gamma}$, $\theta$ and $\eta$ are arbitrary and vanish at the endpoints. Then, this is equivalent to the following equations of motion
\begin{align}\label{michael}
&\dot{\bz} = \Bbb{J}\nabla_{\bz\!}\left\langle\rho | H(\bz)\right\rangle -i\hbar^{-1} \big\langle {Z}\big|\big[ H(\bz),\rho \big] \big\rangle
\,,
\\
&i\hbar\dot{\rho} = \left[ H(\bz),\rho \right]+\nabla_{\bz\!}  \left\langle\rho|H(\bz)\right\rangle\cdot\left[{Z},\rho \right].
\label{frank}
\end{align}
\end{theorem}
\paragraph{Proof.}
This follows by a direct subsistution of the variations in the action principle. We have
\begin{align*}
\delta& \int \bigg( \bzeta(t)\cdot\Bbb{J}\bz(t) +\Big\langle\rho(t),i\hbar\xi(t)+\phi(t)-H(\bz(t))\Big\rangle \bigg) \ \de t= 
\\
=&\int \bigg(
\Bbb{J}\bz\cdot\delta\bzeta 
+\left\langle \rho, \delta\phi \right\rangle
- \left\langle i\hbar\rho,\delta\xi \right\rangle
- \left( \Bbb{J}\bzeta + \left\langle\rho | \nabla_{\bz}H(\bz) \right\rangle \right)\cdot \delta\bz
+\left\langle i\hbar\xi +\phi-H(\bz), \delta\rho \right\rangle
\bigg) \ \de t
\\
=&\int \bigg(
\Bbb{J}\bz\cdot\dot{\boldsymbol{\gamma}}
+\left\langle \rho, \dot{\theta} - \bzeta\cdot\Bbb{J}\boldsymbol{\gamma} \right\rangle
- \left\langle i\hbar\rho,\dot{\eta} - i\hbar^{-1}\big(\left[\xi,{Z}\cdot\Bbb{J}\boldsymbol{\gamma}\right] - \left[ \eta,{Z}\cdot\Bbb{J}\boldsymbol{\zeta} \right]\big) + [\eta,{\xi}] \right\rangle +
\\
&\ - \left( \Bbb{J}\bzeta + \left\langle\rho | \nabla_{\bz}H(\bz) \right\rangle \right)\cdot \left(\boldsymbol{\gamma} - \big\langle [\rho,Z],\eta \big\rangle\right)
+\left\langle i\hbar\xi +\phi-H(\bz), \left[ i\hbar^{-1}{Z}\cdot\Bbb{J}\boldsymbol{\gamma}+ \eta,\rho \right] \right\rangle
 \bigg) \ \de t
\\
=&\int\bigg( \Big\langle 
- \Bbb{J}\dot{\bz} 
- \left\langle \rho | \nabla_{\bz}H(\bz) \right\rangle
+ \left\langle \left[ i\hbar^{-1}\rho,H(\bz) \right], \Bbb{J}Z \right\rangle
, \boldsymbol{\gamma} \Big\rangle 
\\
&\ + \left\langle 
i\hbar\dot{\rho}
+ \left[ \rho, \hat{Z}\cdot  \left\langle \rho | \nabla_{\bz}H(\bz) \right\rangle \right]
+\left[ \rho,H(\bz) \right]
, \boldsymbol{\eta} \right\rangle\bigg) \ \de t
\end{align*}
Then, since $\boldsymbol{\gamma}$, $\theta$ and $\eta$ are arbitrary and vanish at the endpoints, the proof follows. $\quad\blacksquare$

\medskip
\noindent
In order to understand how the above result is related to the usual Ehrenfest equations for quantum expectation dynamics, we immediately observe how these equations  \eqref{michael}-\eqref{frank} are recovered  (along with the evolution of $\rho$) in the case when $\nabla_\bz H(\bz)=0$. As it was pointed out previously, the new feature of equations \eqref{michael}-\eqref{frank} lies in the fact that the expectation values have been considered as independent variables already occurring in the expression of the conserved total energy $\langle H(\bz)\rangle$. This confers the system \eqref{michael}-\eqref{frank} a hybrid classical-quantum structure. Indeed, one observes that new coupled classical-quantum terms  appear in Ehrenfest dynamics: these are the first term on the RHS of \eqref{michael} and the second term on the RHS of \eqref{frank}. 

Notice, the first term on the RHS of  \eqref{michael} does not involve the quantum scales given by $\hbar$. For example, a purely classical system is given by a quantum phase-type Hamiltonian operator of the form  $H(\bz)=h(\bz)\boldsymbol{1}$, where $h(\bz)$ is the classical expression of the Hamiltonian. In this case, while equation \eqref{michael} recovers classical Hamilton's equations, the quantum evolution \eqref{frank} specializes to coherent state dynamics of the type
\[
i\hbar\dot{\rho} = \nabla_\bz h\cdot\left[ {Z},\rho \right].
\]
This establishes how quantum states evolve under the action of purely classical degrees of freedom, thereby enlightening the interplay between classical and quantum dynamics. The same equation can also be obtained by linearizing the  quantum Hamiltonian operator $H(Z)$ around the expectation values (i.e. in the limit $Z\to\bz I$), as prescribed by Littlejohn's nearby orbit approximation for semiclassical mechanics \cite{Littlejohn86}.

\section{Conclusions}

This paper investigated the geometric symmetry properties of quantum and classical-quantum variational principles. Upon departing from the Dirac-Frenkel Lagrangian, different quantum mechanics pictures were recovered from the same variational principle, upon making extensive use of Euler-Poincar\'e theory. This reduction by symmetry naturally leads to consider the Hopf bundle as the natural setting for the Schr\"odinger picture of pure state dynamics, as already proposed by Kibble \cite{kibble1979geometrization}. In addition, new variational principles were presented for mixed state dynamics in both the density matrix and the Wigner-Moyal formulation. Later, new quantum variational principles were also presented for the Heisenberg and Dirac's interaction pictures of quantum dynamics, where the isotropies of the initial state was shown to possess the same role as phases in the Schr\"odinger picture. In particular, Dirac's interaction picture involves the geometric semidirect product structure of two different unitary groups associated to the different quantum propagators arising from the splitting of the Hamiltonian operator.

In the last part of the paper, the Dirac-Frenkel Lagrangian was augmented to account for classical degrees of freedom, incorporating the dynamics of classical motion in hybrid classical-quantum dynamics. As a first step, the mean field model of classical-quantum dynamics was described by using the direct product of the Heisenberg group $\mathcal{H}(\Bbb{R}^{2n})$ (governing classical evolution) and the unitary group $\mathcal{U}(\mathscr{H})$ (governing quantum evolution). Later, Ehrenfest's expectation value dynamics was shown to arise from a novel set of equations, whose dynamics evolves both expectation values and the quantum density matrix under the action of the semidirect product $\mathcal{H}(\Bbb{R}^{2n})\,\circledS\; \mathcal{U}(\mathscr{H})$, whose group and Lie algebra structures were presented. In this context, purely classical dynamics was shown to arise by using Littlejohn's nearby orbit approximation \cite{Littlejohn86}.

\medskip

\subsection*{Acknowledgments} We are indebted with Fran\c{c}ois Gay-Balmaz, Dorje Brody, Jos{\'e} Cari{\~n}ena, Darryl Holm, Alberto Ibort, Henry Jacobs, David Meier, Tudor Ratiu, Paul Skerritt and Alessandro Torrielli  for interesting remarks during the development of this work. Financial support by the Leverhulme Trust Research Project Grant 2014-112, the London Mathematical Society Grant No. 31320 (Applied Geometric Mechanics Network), and the EPSRC Grant No. EP/K503186/1 is greatly acknowledged.

\bigskip

\appendix

\section{Adjoint and coadjoint representations of $\mathcal{H}(\Bbb{R}^{2n})\,\circledS\,\mathcal{U}(\mathscr{H})$.}\label{Appendix1}

First, by using the product rule \eqref{productrule}, one computes the explicit formula for the conjugation action of $\mathcal{H}(\Bbb{R}^{2n})\,\circledS\,\mathcal{U}(\mathscr{H})$ on itself 
\begin{align*}
\text{I}_{(h_1,U_1)} (h_2,U_2) &= (h_1,U_1)(h_2,U_2)(h_1,U_1)^{-1} =
\\
&= \Big( (\mathbf{h}_1, \varphi_1), U_1 \Big) \Big( (\mathbf{h}_2, \varphi_2), U_2 \Big) \Big( (-\mathbf{h}_1, -\varphi_1), U_{h_1}^{\dagger}U^{\dagger}_1U_{h_{1}} \Big) =
\\
&= \Big( (\mathbf{h}_1, \varphi_1), U_1 \Big) \Big( (\mathbf{h}_2-\mathbf{h}_1, \varphi_{2}-\varphi_{1}-\frac12 \mathbf{h}_{2}\cdot\Bbb{J}\mathbf{h}_{1}), U_2\left( U_{h_2} (U_{h_1}^{\dagger}U^{\dagger}_1U_{h_{1}})U_{h_2}^{\dagger} \right) \Big) =
\\
&= \Big( (\mathbf{h}_2, \varphi_2 - \mathbf{h}_2 \cdot\Bbb{J}\mathbf{h}_1),\, U_{1}U_{h_1}U_{2}(U_{h_2} U_{h_1}^{\dagger}U^{\dagger}_1U_{h_{1}}U_{h_2}^{\dagger})U_{h_1}^{\dagger}  \Big)\,.
\end{align*}

\noindent
Then, taking an arbitrary curve 
\[
\left( \mathbf{h}_2(t), \varphi_{2}(t), U_2(t) \right)\in \mathcal{H}(\Bbb{R}^{2n})\,\circledS\,\mathcal{U}(\mathscr{H}) \qquad \text{such that} \quad \left( \mathbf{h}_2(0), \varphi_{2}(0), U_2(0) \right)=\left( 0,0,\text{I} \right)\,, 
\]
and upon denoting $( \dot{\mathbf{h}}_2(0), \dot{\varphi}_{2}(0), \dot{U}_2(0) ) = \left( \boldsymbol{\zeta}, \phi, \xi \right) \in \mathfrak{h}(\Bbb{R}^{2n})\,\circledS\,\mathfrak{u}(\mathscr{H})$ and $\dot{U}_{h_2}(0)=\iota(\zeta)$, one defines the adjoint action of $\mathcal{H}(\Bbb{R}^{2n})\,\circledS\,\mathcal{U}(\mathscr{H})$ on its Lie algebra as follows 
\begin{align*}
\operatorname{Ad}_{(h_1,U_1)}(\zeta,\xi) &= \left.  \frac{\de}{\de t} \right\vert_{t=0} \text{I}_{(h_1,U_1)} (h_2(t),U_2(t)) = 
\\
&=  \left.  \frac{\de}{\de t} \right\vert_{t=0} \Big(  (\mathbf{h}_2(t), \varphi_2(t) - \mathbf{h}_2(t) \cdot\Bbb{J}\mathbf{h}_1),\, U_{1}U_{h_1}U_{2}(t)(U_{h_2}(t) U_{h_1}^{\dagger}U^{\dagger}_1U_{h_{1}}U_{h_2}^{\dagger}(t))U_{h_1}^{\dagger}  \Big) =
\\
&= \Big(  (\dot{\mathbf{h}}_2(0), \dot{\varphi}_2(0) - \dot{\mathbf{h}}_2(0) \cdot\Bbb{J}\mathbf{h}_1),\, U_{1}U_{h_1}\dot{U}_{2}(0)(U_{h_2}(0) U_{h_1}^{\dagger}U^{\dagger}_1U_{h_{1}}U_{h_2}^{\dagger}(0))U_{h_1}^{\dagger} + 
\\ 
&\quad+ U_{1}U_{h_1}U_{2}(0)(\dot{U}_{h_2}(0) U_{h_1}^{\dagger}U^{\dagger}_1U_{h_{1}}U_{h_2}^{\dagger}(0))U_{h_1}^{\dagger} +
\\
&\quad - U_{1}U_{h_1}U_{2}(0)(U_{h_2}(0) U_{h_1}^{\dagger}U^{\dagger}_1U_{h_{1}}U_{h_2}^{\dagger}(0)\dot{U}_{h_2}(0)U_{h_2}^{\dagger}(0))U_{h_1}^{\dagger}  \Big) =
\\
&= \Big( (\boldsymbol{\zeta},\varphi-\boldsymbol{\zeta}\cdot\Bbb{J}\mathbf{h}_1),\, U_1U_{h_1}(\xi+\iota(\zeta))U_{h_1}^{\dagger}U_1^{\dagger} - U_{h_1}\iota(\zeta)U_{h_1}^{\dagger} \Big) =
\\
&= \Big( \operatorname{Ad}_{h}\zeta, \,U_1U_{h_1}(\xi+\iota(\zeta))U_{h_1}^{\dagger}U_1^{\dagger} - \iota(\operatorname{Ad}_{h}\zeta )  \Big)\,.
\end{align*}

\noindent
At this point, using the notation $(\nu,\mu)=((\boldsymbol\nu,\alpha),\mu) \in \mathfrak{h}(\Bbb{R}^{2n})^*\times\mathfrak{u}(\mathscr{H})^*\simeq\Bbb{R}^{2n+1}\times\mathfrak{u}(\mathscr{H})^*$, one computes the coadjoint representation on $\mathcal{H}(\Bbb{R}^{2n})\,\circledS\,\mathcal{U}(\mathscr{H})$ via the pairing 
\begin{align*}
\Big\langle  \operatorname{Ad}^*_{(h,U)}(\nu,\mu), \, (\zeta,\xi) \Big\rangle &= \Big\langle  (\nu,\mu), \, \operatorname{Ad}_{(h,U)} (\zeta,\xi) \Big\rangle = 
\\
&= \Big\langle  (\nu,\mu), \,  \left(\operatorname{Ad}_{h}\zeta, \,U_1U_{h_1}(\xi+\iota(\zeta))U_{h_1}^{\dagger}U_1^{\dagger} - U_{h_1}\iota(\zeta)U_{h_1}^{\dagger} \right)\Big\rangle = 
\\
&= \Big\langle \Big(\operatorname{Ad}^*_{h}\nu + \iota^*\left( U_h^{\dagger}(U^{\dagger}\mu U -\mu)U_h\right), U_h^{\dagger}U^{\dagger}\mu U_h U_h \Big),\, (\zeta,\xi)  \Big\rangle\,,
\end{align*}
where $\operatorname{Ad}^*_{h}\nu = \left( \boldsymbol{\nu} - \alpha\Bbb{J}\mathbf{h}, \alpha \right)$ is the coadjoint representation on $\mathcal{H}(\Bbb{R}^{2n})$, and $\iota^*:\mathfrak{u}^*(\mathscr{H})\to\mathfrak{h}^*(\Bbb{R}^{2n})$ is the dual of the Lie algebra homomorphism generated by the displacement operator $U_h$, which is also computed via the pairing giving the following expression 
\[
\iota^*(\mu) = \Big( \langle \mu, -i\hbar^{-1}\Bbb{J}Z \rangle, \text{Tr}(i\hbar^{-1}\mu) \Big)
\,,\qquad
\qquad
\forall
\mu\in\mathfrak{u}^*(\mathscr{H})\,.
\]
By direct substitution, the coadjoint action on $\mathcal{H}(\Bbb{R}^{2n})\,\circledS\,\mathcal{U}(\mathscr{H})$ reads
\[
\operatorname{Ad}^*_{(h,U)}(\nu,\mu)=\Big(\boldsymbol\nu-\alpha\Bbb{J}\mathbf{h}+\big\langle\mu-U^\dagger\mu U,i\hbar^{-1}\Bbb{J}Z\big\rangle, \, \alpha,\,U_h^\dagger U^\dagger\mu UU_h\Big)\,.
\]

\bigskip




\begin{thebibliography}{10}

\bibitem{AhAn87}
Yakir Aharonov and J~Anandan.
\newblock Phase change during a cyclic quantum evolution.
\newblock {\em Physical Review Letters}, 58(16):1593, 1987.

\bibitem{Anandan88non}
J~Anandan.
\newblock Non-adiabatic non-abelian geometric phase.
\newblock {\em Physics Letters A}, 133(4):171--175, 1988.

\bibitem{Anandan91}
J.~Anandan.
\newblock A geometric approach to quantum mechanics.
\newblock {\em Foundations of Physics}, 21(11):1265--1284, 1991.

\bibitem{AnAh90}
J.~Anandan and Y.~Aharonov.
\newblock Geometry of quantum evolution.
\newblock {\em Phys. Rev. Lett.}, 65:1697--1700, Oct 1990.

\bibitem{andersson2013dynamic}
Ole Andersson and Hoshang Heydari.
\newblock Dynamic distance measure on spaces of isospectral mixed quantum
  states.
\newblock {\em Entropy}, 15(9):3688--3697, 2013.

\bibitem{AshAb95}
Abhay Ashtekar and Troy~A. Schilling.
\newblock Geometry of quantum mechanics.
\newblock {\em AIP Conference Proceedings}, 342(1), 1995.

\bibitem{BoCaGra91}
Luis~J Boya, Jos{\'e}F Cari{\~n}ena, and Jos{\'e}M Gracia-Bond{\'\i}a.
\newblock Symplectic structure of the aharonov-anandan geometric phase.
\newblock {\em Physics Letters A}, 161(1):30--34, 1991.

\bibitem{BroHu01}
Dorje~C. Brody and Lane~P. Hughston.
\newblock Geometric quantum mechanics.
\newblock {\em Journal of Geometry and Physics}, 38(1):19 -- 53, 2001.

\bibitem{CarPa85}
Richard Car and Mark Parrinello.
\newblock Unified approach for molecular dynamics and density-functional
  theory.
\newblock {\em Physical review letters}, 55(22):2471, 1985.

\bibitem{CarHoKoOku06}
Alberto Carlini, Akio Hosoya, Tatsuhiko Koike, and Yosuke Okudaira.
\newblock Time-optimal quantum evolution.
\newblock {\em Phys. Rev. Lett.}, 96:060503, Feb 2006.

\bibitem{CarHoKoOku07}
Alberto Carlini, Akio Hosoya, Tatsuhiko Koike, and Yosuke Okudaira.
\newblock Time-optimal unitary operations.
\newblock {\em Phys. Rev. A}, 75:042308, Apr 2007.

\bibitem{cendra2001lagrangian}
Hern{\'a}n Cendra, Jerrold~E Marsden, and Tudor~S Ratiu.
\newblock {\em Lagrangian reduction by stages}, volume 722.
\newblock American Mathematical Soc., 2001.

\bibitem{ChiMel2012}
Eduardo Chiumiento and Michael Melgaard.
\newblock Stiefel and grassmann manifolds in quantum chemistry.
\newblock {\em Journal of Geometry and Physics}, 62(8):1866--1881, 2012.

\bibitem{Chrus94}
Dariusz Chru{\'s}ci{\'n}ski.
\newblock Symplectic structure for the non-abelian geometric phase.
\newblock {\em Physics Letters A}, 186(1):1--4, 1994.

\bibitem{CleMar08}
J.~{Clemente-Gallardo} and G.~{Marmo}.
\newblock {Basics of Quantum Mechanics, Geometrization and Some Applications to
  Quantum Information}.
\newblock {\em International Journal of Geometric Methods in Modern Physics},
  5:989, 2008.

\bibitem{Dalessandro01}
D.~D'Alessandro and M.~Dahleh.
\newblock Optimal control of two-level quantum systems.
\newblock {\em Automatic Control, IEEE Transactions on}, 46(6):866--876, Jun
  2001.

\bibitem{dirac1930note}
Paul~AM Dirac.
\newblock Note on exchange phenomena in the thomas atom.
\newblock In {\em Mathematical Proceedings of the Cambridge Philosophical
  Society}, volume~26, pages 376--385. Cambridge Univ Press, 1930.

\bibitem{FaKuMaMa10}
Paolo Facchi, Ravi Kulkarni, V.I. Man'ko, Giuseppe Marmo, E.C.G. Sudarshan, and
  Franco Ventriglia.
\newblock Classical and quantum fisher information in the geometrical
  formulation of quantum mechanics.
\newblock {\em Physics Letters A}, 374(48):4801 -- 4803, 2010.

\bibitem{frenkel1934wave}
Yakovich~Ilich Frenkel.
\newblock Wave mechanics.
\newblock 1934.

\bibitem{gay2010reduction}
Fran{\c{c}}ois Gay-Balmaz and Cesare Tronci.
\newblock Reduction theory for symmetry breaking with applications to nematic
  systems.
\newblock {\em Physica D: Nonlinear Phenomena}, 239(20):1929--1947, 2010.

\bibitem{Grig92}
A.~N. Grigorenko.
\newblock Geometry of projective hilbert space.
\newblock {\em Phys. Rev. A}, 46:7292--7294, Dec 1992.

\bibitem{GunWaNa2014}
Utkan G{\"u}ng{\"o}rd{\"u}, Yidun Wan, and Mikio Nakahara.
\newblock Non-adiabatic universal holonomic quantum gates based on abelian
  holonomies.
\newblock {\em Journal of the Physical Society of Japan}, 83(3), 2014.

\bibitem{Heller76}
Eric~J Heller.
\newblock Time dependent variational approach to semiclassical dynamics.
\newblock {\em The Journal of Chemical Physics}, 64(1):63--73, 1976.

\bibitem{holm1998euler}
Darryl~D Holm, Jerrold~E Marsden, and Tudor~S Ratiu.
\newblock The euler--poincar{\'e} equations and semidirect products with
  applications to continuum theories.
\newblock {\em Advances in Mathematics}, 137(1):1--81, 1998.

\bibitem{holm2009geometric}
Darryl~D Holm, Tanya Schmah, Cristina Stoica, and David~CP Ellis.
\newblock {\em Geometric mechanics and symmetry: from finite to infinite
  dimensions}.
\newblock Number~12. Oxford University Press London, 2009.

\bibitem{KhaBroGla01}
Navin Khaneja, Roger Brockett, and Steffen~J. Glaser.
\newblock Time optimal control in spin systems.
\newblock {\em Phys. Rev. A}, 63:032308, Feb 2001.

\bibitem{kibble1979geometrization}
TWB Kibble.
\newblock Geometrization of quantum mechanics.
\newblock {\em Communications in Mathematical Physics}, 65(2):189--201, 1979.

\bibitem{kramer1981geometry}
PH~Kramer and Marcos Saraceno.
\newblock {\em Geometry of the time-dependent variational principle in quantum
  mechanics}.
\newblock Springer, 1981.

\bibitem{Littlejohn86}
Robert~G Littlejohn.
\newblock The semiclassical evolution of wave packets.
\newblock {\em Physics reports}, 138(4):193--291, 1986.

\bibitem{LowMuk72}
P-O L{\"o}wdin and PK~Mukherjee.
\newblock Some comments on the time-dependent variation principle.
\newblock {\em Chemical Physics Letters}, 14(1):1--7, 1972.

\bibitem{Lucarelli2005}
Dennis Lucarelli.
\newblock Control aspects of holonomic quantum computation.
\newblock {\em Journal of mathematical physics}, 46(5):052103, 2005.

\bibitem{marsden1999introduction}
Jerrold~E Marsden and Tudor~S Ratiu.
\newblock {\em Introduction to mechanics and symmetry: a basic exposition of
  classical mechanical systems}, volume~17.
\newblock Springer, 1999.

\bibitem{MiWa01}
Akimasa Miyake and Miki Wadati.
\newblock Geometric strategy for the optimal quantum search.
\newblock {\em Phys. Rev. A}, 64:042317, Sep 2001.

\bibitem{Montgomery91}
R~Montgomery.
\newblock Heisenberg and isoholonomic inequalities.
\newblock {\em Symplectic Geometry and Mathematical Physics, Editors P. Donato
  et. al., Birkh{\"a}user, Boston}, pages 303--325, 1991.

\bibitem{moyal1949quantum}
Jos{\'e}~Enrique Moyal.
\newblock Quantum mechanics as a statistical theory.
\newblock In {\em Mathematical Proceedings of the Cambridge Philosophical
  Society}, volume~45, pages 99--124. Cambridge Univ Press, 1949.

\bibitem{Ohta2000}
Katsuhisa Ohta.
\newblock Time-dependent variational principle with constraints.
\newblock {\em Chemical Physics Letters}, 329(3):248--254, 2000.

\bibitem{Poulsen11}
Jens~Aage Poulsen.
\newblock A variational principle in wigner phase-space with applications to
  statistical mechanics.
\newblock {\em The Journal of Chemical Physics}, 134(3):--, 2011.

\bibitem{Salcedo2012}
LL~Salcedo.
\newblock Statistical consistency of quantum-classical hybrids.
\newblock {\em Physical Review A}, 85(2):022127, 2012.

\bibitem{SaHuKu2011}
Adam Sawicki, Alan Huckleberry, and Marek Ku{\'s}.
\newblock Symplectic geometry of entanglement.
\newblock {\em Communications in Mathematical Physics}, 305(2):441--468, 2011.

\bibitem{TaNaHa2005}
Shogo Tanimura, Mikio Nakahara, and Daisuke Hayashi.
\newblock Exact solutions of the isoholonomic problem and the optimal control
  problem in holonomic quantum computation.
\newblock {\em Journal of mathematical physics}, 46(2):022101, 2005.

\bibitem{Wyatt2006}
C.J. Trahan and R.E. Wyatt.
\newblock {\em Quantum Dynamics with Trajectories: Introduction to Quantum
  Hydrodynamics}.
\newblock Interdisciplinary Applied Mathematics. Springer, 2006.

\bibitem{uhlmann2009geometry}
Armin Uhlmann and Bernd Crell.
\newblock Geometry of state spaces.
\newblock In {\em Entanglement and Decoherence}, pages 1--60. Springer, 2009.

\bibitem{wigner1932quantum}
Eugene Wigner.
\newblock On the quantum correction for thermodynamic equilibrium.
\newblock {\em Physical Review}, 40(5):749, 1932.

\bibitem{Zachos2005}
Cosmas Zachos, David Fairlie, and Thomas Curtright.
\newblock {\em Quantum mechanics in phase space: an overview with selected
  papers}, volume~34.
\newblock World Scientific, 2005.

\bibitem{Zhao2012}
Lian-Jie Zhao, Yan-Song Li, Liang Hao, Tao Zhou, and Gui~Lu Long.
\newblock Geometric pictures for quantum search algorithms.
\newblock {\em Quantum Information Processing}, 11(2):325--340, 2012.

\end{thebibliography}
\end{document}